\documentclass[acmsmall]{acmart}
\setcopyright{none}
\acmDOI{}
\acmPrice{}
\acmISBN{}

\usepackage{url}
\usepackage{cprotect} 
\usepackage{fancyvrb} 
\usepackage[super]{nth}
\usepackage[caption=false,font=footnotesize]{subfig}
\usepackage[linesnumbered,ruled]{algorithm2e}
\usepackage[para,online,flushleft]{threeparttable}

\def\ie{{i.e.},~}
\def\eg{{e.g.},~}

\let\oldnl\nl
\newcommand{\nonl}{\renewcommand{\nl}{\let\nl\oldnl}}

\newlength\mylen
\newcommand\myinput[1]{%
  \settowidth\mylen{\KwIn{}}%
  \setlength\hangindent{\mylen}%
  \hspace*{\mylen}#1\\}

\DeclareMathOperator*{\argmin}{arg\,min}

\title{Adversarial Planning}

\author{Valentin Vie}
\email{vav4@cse.psu.edu}
\affiliation{%
  \institution{Pennsylvania State University}
  \streetaddress{Westgate Building, W378}
  \city{University Park}
  \state{Pennsylvania}
  \country{USA}
  \postcode{16801}
}
\author{Ryan Sheatsley}
\email{sheatsley@psu.edu}
\affiliation{%
  \institution{Pennsylvania State University}
  \country{USA}
}
\author{Sophia Beyda}
\email{beyda@psu.edu}
\affiliation{%
  \institution{Pennsylvania State University}
  \country{USA}
}
\author{Sushrut Shringarputale}
\email{sps5394@psu.edu}
\affiliation{%
  \institution{Pennsylvania State University}
  \country{USA}
}
\author{Kevin Chan}
\email{kevin.s.chan.civ@mail.mil}
\affiliation{%
  \institution{Army Research Laboratory}
  \country{USA}
}
\author{Trent Jaeger}
\email{tjaeger@cse.psu.edu}
\affiliation{%
  \institution{Pennsylvania State University}
  \country{USA}
}
\author{Patrick McDaniel}
\email{mcdaniel@cse.psu.edu}
\affiliation{%
  \institution{Pennsylvania State University}
  \country{USA}
}

\renewcommand{\shortauthors}{Vie et al.}

\begin{document}

\begin{abstract}
Planning algorithms are used in computational systems to direct autonomous behavior.
In a canonical application for example, planning for autonomous vehicles is used to automate the static or continuous planning towards performance, resource management, or functional goals (e.g., arriving at the destination, managing fuel consumption).
Existing planning algorithms assume non-adversarial settings; a least cost plan is developed based on available environmental information (i.e., the input instance).  Yet, it is unclear how such algorithms will perform in the face of adversaries attempting to thwart the planner.  In this paper, we explore the security of planning algorithms as used in cyber- and cyber-physical systems.  We present two {\it adversarial planning} algorithms--one static and one adaptive--that perturb input planning instances to maximize cost (often substantially so).  We evaluate the performance of the algorithms against two dominant planning algorithms used in commercial applications (D* Lite and Fast Downward) and show both are vulnerable to extremely limited adversarial action.  Here, experiments show that the adversary is able to increase plan costs in 66.9\% of instances by only removing a single action from the actions space (D* Lite) and render 70\% of the instances from an international planning competition unsolvable by removing only three actions (Fast Forward).  Finally, we show that finding an optimal perturbation in any search-based planning system is NP-hard.
\end{abstract}

\maketitle

\section{Introduction}

The science of planning is about creating a plan---a set of actions---to achieve a goal. Applications of planning algorithms are found in robotics, aerospace, and industrial processes, where they are used to find the most optimal solution to a given problem. For example, planning algorithms have been used for unmanned vehicles~\cite{DARPA}, natural language generation~\cite{natural_language}, greenhouse logistics~\cite{smart_greenhouses},  manufacturing~\cite{bsnw80}, network vulnerability analysis~\cite{cyberPlanning}, and navigation~\cite{D*Lite}.

\begin{figure}[!t]
\centerline{\includegraphics[width=1\linewidth]{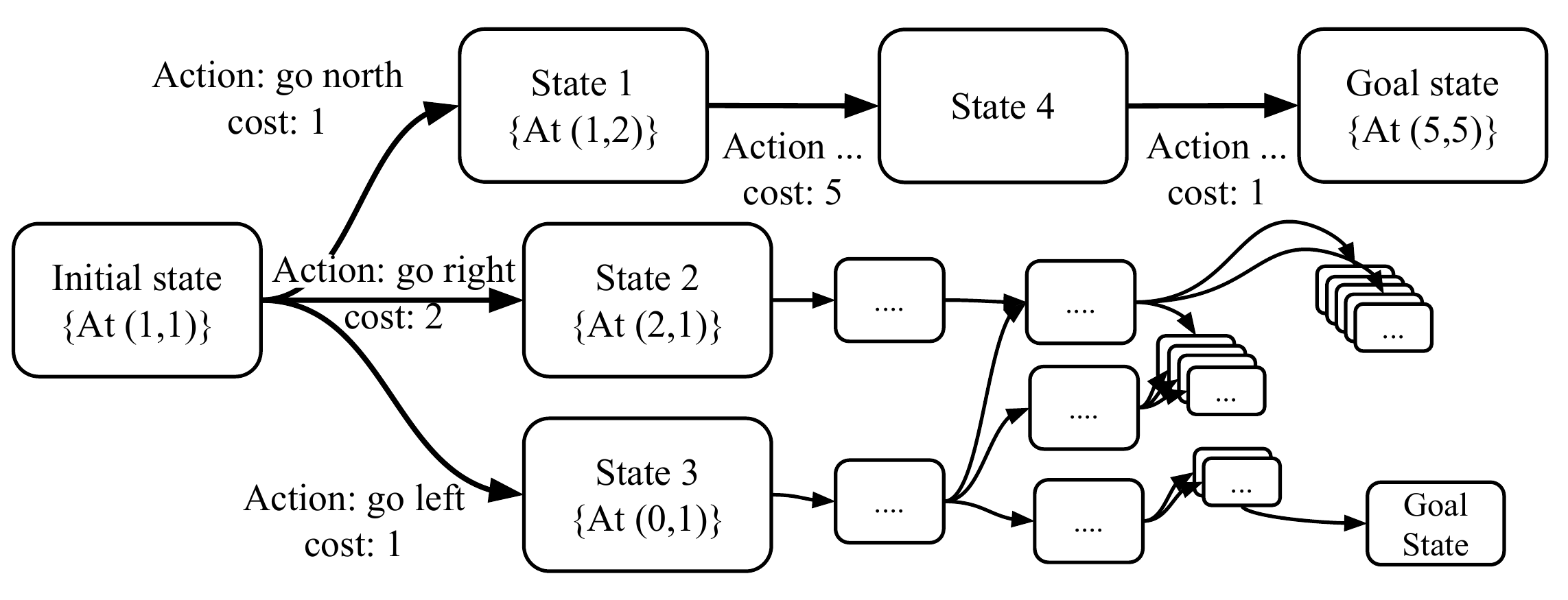}}
\caption{State-space viewed as a graph with actions linking the different states. A planner finds a way to reach a goal state
}
\label{HSP_example}
\end{figure}

Plans are, in most instances, sequences of steps called \textit{actions}. Each action represents an atomic operation that an agent performs to achieve some sub-goal within the domain, \eg moving a step forward or  loading or unloading a device. A planning algorithm receives an input planning problem consisting of states and operations and outputs a sequence of actions, which, when executed from the initial state, gets the agent to a goal state. The \textit{plan cost} is the sum of the costs of the actions in the plan. Planning algorithms are designed to minimize the plan cost (which can include multiple metrics such as distance traveled, resources, etc.). A common visualization is to create a graph with states as nodes and actions as edges connecting two states (Fig.~\ref{HSP_example}).

In practice, planners (often called agents) are embedded in larger systems of components including sensors (e.g., motion detection, LIDAR), classification or logic systems (e.g., machine learning-based image recognition, monotonic reasoning), logic/software driven mechanical actuators (e.g., break systems, batteries), and embedded operating systems~\cite{bt15,zzwzc18}.  For example, autonomous vehicles simultaneously use local and global planners to make short-term (e.g., motion planning through a busy intersection) and long term (e.g., route planning) decisions on how to safely direct the vehicle to its destination~\cite{sar18}.  While the security of many of these components and the system as a whole have been explored in many contexts~\cite{tw16,arc+15}, few, if any, previous efforts have attempted to understand how these systems can be subverted by an adversary attacking the planning process.

This work considers an adversary attempting to subvert a system by attacking the planner with the goal of reducing the effectiveness of the system (e.g., inducing long indirect routing) or in the degenerate case preventing the successful completion of the plan entirely (e.g., preventing the vehicle from reaching its destination).  Here, we explore a threat model in which an adversary is able to remove a number of actions from the set of actions in the action space available to the agent.   We consider two adversarial strategies: one where the adversary perturbs the elements of the input planning instance (called a static or {\it offline} attack) and another where the adversary adaptively counteracts an executing plan (called an {\it online} attack).  Here we build upon security efforts at identifying worst case inputs to complex systems and algorithms, e.g., in adversarial machine learning~\cite{carlini2017towards,Papernot,kurakin2016adversarial} or fuzzing~\cite{chen2018angora,aschermann2019redqueen}.

The consequences of inefficient or unachievable plans (the outcome of an attack) can be dire.  For example, a poorly designed motion plan in an autonomous vehicle can lead to unsafe conditions or accidents~\cite{pcz+16}.  Poor or sub-optimal planning in chemical manufacturing (called chemical production scheduling) can lead to production quality problems, induce equipment failures, or reduce efficiency~\cite{Kidam2013AnalysisOE}.  Planning failures in transportation systems or logistics can cause widespread outages and lead to travelers being stranded~\cite{ws19} (see Section~\ref{sec:fdown} in which we simulate a realistic attack on the Munich airport planning systems).  In short, we posit any system that depends on a viable and efficient plan can be undermined by an adversary with the ability to make small perturbations to the plan space.  Similar to work in adversarial machine learning, the adversary's goal to (a) find a perturbation that achieves the negative outcome (increasing plan execution cost) while (b) minimizing the size of the perturbation (number of actions removed).

In this paper, we develop and evaluate two adversarial algorithms that manipulate real-world planning systems to induce sub-optimal (i.e., more costly) plans.  We introduce the {\it window-heuristic} as an approximation function to predict the expected plan cost impact made by removing part of the action space (called an adversarial perturbation of the input planning instance).  Several adversarial algorithms are presented, and intuition and examples are provided.

We empirically evaluate our approach on the D* Lite algorithm~\cite{D*Lite} (a path-finding algorithm) and the Fast Downward planner~\cite{FastDownward}, two of the most widely used planning algorithms used on industrial systems.   The D* Lite algorithm has been widely used for autonomous vehicle navigation for its capacity to adapt to changes in the environment. The original version (D*) was, for example, used by DARPA for its Unmanned Ground Vehicle program~\cite{DARPA} and for Mars rover prototypes~\cite{D*Lite}. The Fast Downward planner is a classical planning system based on {\it forward heuristic search}~\cite{FastDownward} and was used by two winning planners of the sequential-satisfying track to find low-cost plans in a fixed amount of time of the \nth{9} International Planning Competition (IPC).

Our experiments of D* Lite planning show that 66.9\% of randomly selected planning instances in a size $15\times15$ maze\footnote{For reference, routing in this grid is equivalent to real world taxi route planning during periods of high congestion (e.g., New Years eve) in mid-town New York City (i.e., bounded by 8th Avenue (W), 59th (N), the East River (E), and Times Square(S)).} have an increased cost or become unsolvable with a single perturbation (82.7\% with two). Unsolvable is defined as the inability for the agent to find a valid path to the goal state. In a second set of experiments, we evaluate the performance of the window-heuristic on the international 2014 IPC competition planning instances.  Interestingly, for some competition domains, we found that 70\% of instances become unsolvable if an adversary can remove only three actions out of over 500 available.

\vspace{3pt}
\noindent
We make the following contributions in this work:
\begin{itemize}

    \item We develop an algorithm to find adversarial changes in STRIPS-written tasks and path-finding problems. We use our window-heuristic to create a table of adversarial changes which we use to perturb an planning instance (Section~\ref{Approach}).

    \item The attacks are applied to two of the dominant planning systems used in commercial applications: D* Lite and the Fast Downward planner in real-world settings (Section~\ref{Evaluation}).


    \item The online attack achieved an 82\% success rate at inducing cost while the offline attack reached up to a 100\% success rate, depending upon the domain considered.

\end{itemize}

Section~\ref{sec:realizing} further presents the high-level results of a extended survey of the security of fielded planning systems.  In this, we explore the realism of the proposed threat model and provide concrete examples of how (and why) such perturbations can be achieved in real systems and applications such as autonomous vehicles, manufacturing, and data center management.

\section{Background}

\begin{figure}[!t]
\centerline{\includegraphics[width=0.8\linewidth]{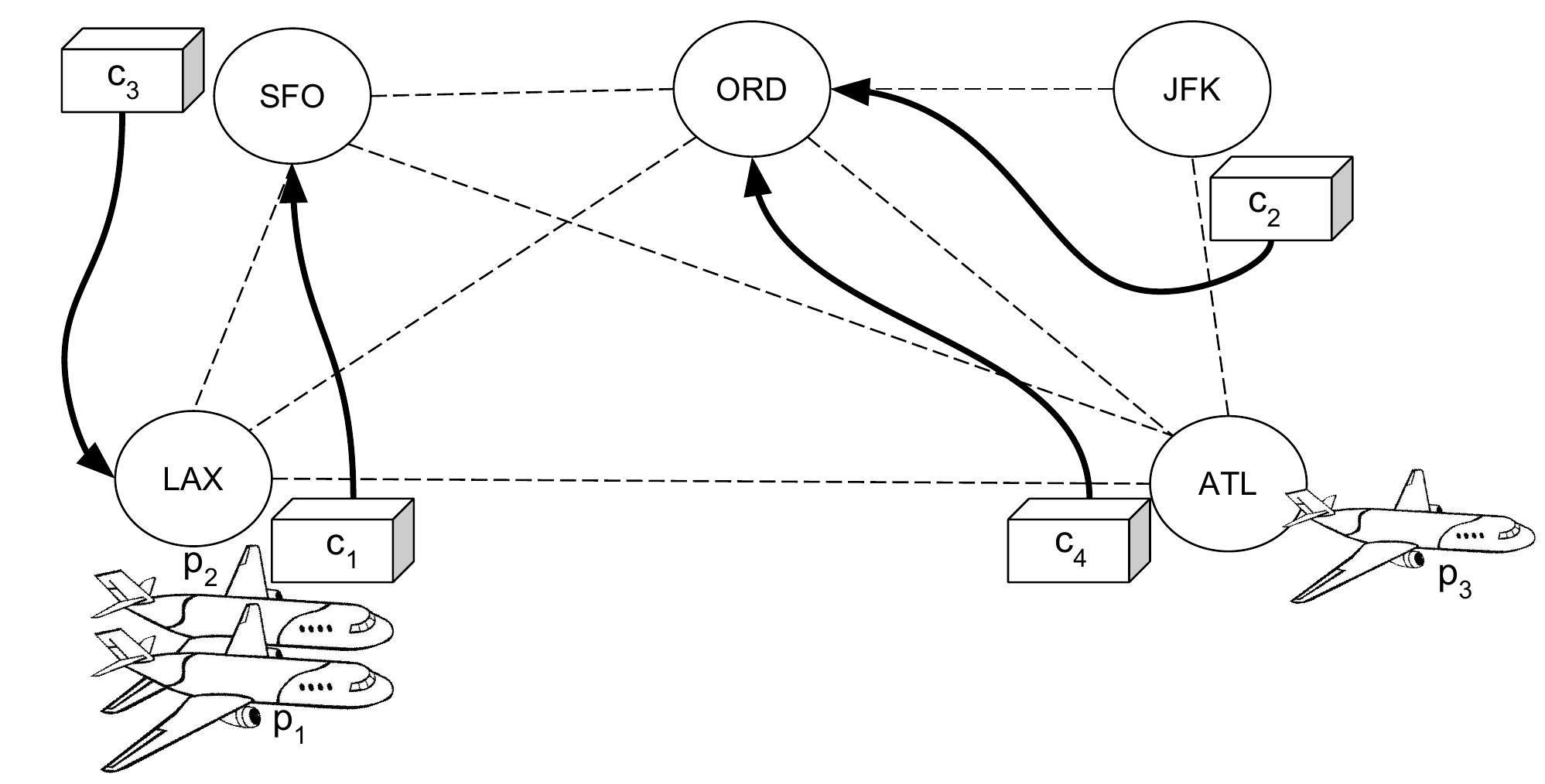}}
\cprotect\caption{{\small \verb|air cargo transportation|}: planning to deliver the packages to the destination with the minimum number of actions ({\small \verb|Load|}, {\small \verb|Unload|}, {\small \verb|Fly|}).}
\label{fig:air_cargo_pb}
\end{figure}

\textbf{Planning Algorithms} - The objective of a planning algorithm is to output a plan containing different actions to achieve a goal. This objective can take different forms depending on the problem. For example, given a Rubik's cube, the goal is to have one color on each face. For the {\small \verb|air cargo transportation|} domain, the goal is to deliver all packages to their destination airports (Fig.~\ref{fig:air_cargo_pb}). A planning algorithm outputs a sequence of actions $A_1, A_2,...,A_n$ which, when executed from the initial state $S_{init}$, gets the agent to a goal state $S_{goal}$. A \textit{state} is defined as the configuration of the environment. There can be as many states as the environment requires. For example, a Rubik's cube contains more than 43 quintillion different states.
If we call $f$ the transition function, then we have the following state trajectory: $S_{init}=S_{0}, S_1 = f(S_{init},A_1), ..., S_{goal} = f(S_{n-1},A_{n}) = S_{n}$. The solution is said to be optimal if the total cost $\sum_{i=0}^{n-1} c(S_i, A_{i+1})$ is minimized, where $c(S_i, A_{i+1})$ is the cost of action $A_{i+1}$ from state $S_i$. The cost function is chosen by the user depending on the criterion under optimization, e.g., time, resources, etc.

We distinguish two categories of planning systems: online and offline. We define a planning system as offline when the state trajectory to the goal state is computed before the execution begins. The entire plan is computed beforehand, which leaves little room for adaptation but leaves enough time to optimize the solution. On the other hand, with an online planner, an agent can react to changes in the environment (miscalculation, error, obstacle discovery, etc.). The plan is updated as the agent moves. Examples of these types of planning systems are the Mars Rover's navigation (offline) and a Tesla's navigation system (online).

\begin{figure}[!t]
{\small
\begin{Verbatim}[frame=single]
(:action LOAD
  :parameters (?c - cargo ?p - plane
               ?a - airport)
  :precondition (and (At ?c ?a) (At ?p ?a))
  :effect (and (In ?c ?p) (not (At ?c ?a))))
\end{Verbatim}
}
{\small
\begin{Verbatim}[frame=single]
(:action LOAD_c1_p1_LAX
  :parameters (c1 - cargo p1 - plane
               LAX - airport)
  :precondition (and (At c1 LAX) (At p1 LAX))
  :effect (and (In c1 p1) (not (At c1 LAX))))
\end{Verbatim}
}
\cprotect\caption{(Top) Non-grounded {\small \verb|Load|} action for {\small \verb|air cargo transportation|} domain. (Bottom) A grounded {\small \verb|Load|} operator.
}
\label{fig:LoadSTRIPS}
\end{figure}

\vspace{3pt}\noindent\textbf{STRIPS Notation} - STRIPS is a standard language to describe classical planning problems. Simply put, a planning problem articulated in STRIPS is a triple $(S_{init}, S_{goal}, \mathcal{O})$. See Appendix~\ref{Aircargodomain_example} for an example of a STRIPS instance.

The first item $S_{init}$ is the initial state represented as a collection of variables known as \textit{atoms} or \textit{predicates}. For example, to specify that {\small \verb|cargo1|} is at {\small \verb|LAX|} airport (Fig.~\ref{fig:air_cargo_pb}), we would describe the state with the predicate {\small \verb|(At, cargo1, LAX)|}.  The second element $S_{goal}$ is a characterization of the goal state, defined as a set of predicates.   In {\small \verb|air cargo transportation|} goal is the final position of the cargo.
The last item $\mathcal{O}$ is a finite set of grounded actions.  A grounded action is an action in which all parameters have been assigned to real objects, i.e., the action has been fully specified.  A non-grounded action has at least one parameter not bound to a specific object. We call \textit{actions} operators that can transform a state into (potentially) another state by impacting the cost of the plan. Only deterministic actions are considered, meaning there is no uncertainty about the outcome of an action. Three sets of predicates define an operator: parameters, preconditions, and effects. The parameters specify the objects' type. As demonstrated in Figure~\ref{fig:LoadSTRIPS}, {\small \verb|c|} has to be cargo, {\small \verb|p|} a plane, and {\small \verb|a|} an airport. The preconditions must be satisfied for the action to be executed. For example, in Figure~\ref{fig:LoadSTRIPS}, in order to load a unit of cargo into a plane, the cargo must be at the same airport as the plane. The effects impact the input state modifying its predicates, they define the consequences of the action on the state. The set $\mathcal{O}$ can contain thousands of grounded operators. For example, in Figure~\ref{fig:air_cargo_pb}, there would be one grounded {\small \verb|Load|} operator for each combination of cargo, plane, and airport. Consequently, we mostly use non-grounded actions to define a planning task. For example, in Figure~\ref{fig:LoadSTRIPS}, {\small \verb|a|} could be any airport (SFO, LAX, etc.). If two planning problems share the same non-grounded actions, they belong to the same \textit{domain}, e.g., instances of the Rubik's cube have different starting positions, but are always solved by rotating the faces (\ie the same operators).

Finally, within the scope of this paper, we introduce an adversarial change as a grounded operator being removed from the set of operators $\mathcal{O}$, with the goal of increasing the cost. The goal for the adversary is to find $k$ grounded actions to remove from $\mathcal{O}$ to increase the cost of the plan. In the offline attack, the removal happens before the agent starts planning whereas in the online attack, the removal happens while the agent is executing the plan. We consider $k$ to never exceed ten and $|\mathcal{O}|$ can have an arbitrary size (usually thousands of grounded actions).

\vspace{3pt}\noindent\textbf{General Purpose Planners} - Since we develop a general heuristic to find effective adversarial changes in any domain, we work with domain-independent planners. General purpose planners generally perform worse than domain-specific ones because they lack the domain-specific knowledge to prune the search for a solution~\cite{Sokoban_Domain_Specific}.  However, they offer a convenient way to solve planning tasks without domain---and planning---expertise. These planners are divided into three main categories~\cite{Russell}. Currently, the most popular and effective approaches to solving a deterministic planning problem are: (1) performing a search using a planning graph~\cite{GraphPlan}, (2) translating it to a Boolean satisfiability (SAT) problem and using a SAT solver~\cite{blackboxSAT}, or (3) executing a forward or backward search in the state-space with a heuristic~\cite{Bonet01planningas}. We build the window-heuristic (Section~\ref{Approach}) on top of the last category of planners: forward heuristic search planners. We also test the attack on forward heuristic search planners although our techniques apply to other classes of planners as well.

In more detail, the planners we focus on here perform a forward or backward search in the state-space: a graph where the nodes are the states and the directed edges are the grounded actions (See Fig.~\ref{HSP_example}). The goal is to find a minimum cost path from the initial state to the goal state in the state-space. The search is guided by a heuristic by estimating the distance from a particular state to the goal state. Heuristics can be extracted from the problem directly or can be specific to a domain. For path-finding problems, a common practice is to use the Manhattan distance or the Euclidean distance. The competition planner Fast Downward~\cite{participatingPlanners} can run with different search algorithms (\eg A* and Best-first search) and different heuristics (\eg FF heuristic and Additive heuristic). While this work focuses on forward/backward search, our preliminary analysis of other general-purpose planners suggests vulnerability to adversarial manipulation is a function of the instance and less on the specific planning algorithm.  We defer that analysis to future work.

\section{Threat Model}
\label{Adversarial_goals}
We develop an algorithm to adversely influence planning systems. Given a planning task, we want to output a set of adversarial changes to decrease the cost of the initial plan. The metric used to measure the cost of the plan changes between the different applications of planning systems. It can be the computation time, the algorithmic complexity, the resource requirements, or the cost function. For instance, the complexity is critical for online planning algorithms such as on-board planners in unmanned vehicles, where the energy resources are limited~\cite{TREX}.

Specifically, an adversary will seek to come up with adversarial changes that will impact the cost of the plan. The computation time needed to find a plan may also be affected by the adversarial changes. Indeed, a planning system has to potentially explore deeper in the state-space to find an acceptable solution. The ultimate goal for the adversary is to make sure the agent will never reach the goal state without the agent knowing it. In this way, the cost of the plan is infinite, and the planner can loop almost forever. Note that perturbing a plan is not always feasible; it depends on the adversary's capabilities and the instance considered.

\subsection{Adversarial Capabilities}
\label{Adversarial_capabilities}

The strength of an adversary is defined by the information and capabilities at their disposal. We consider two kinds of attacks: online and offline.  In the offline case, the adversary perturbs the input planning instance given to the agent's planner.
For example, given a task in the {\small \verb|air cargo| \verb|transportation|} domain, the agent needs to find a series of actions reaching the goal, without {\small \verb|(Load cargo1 in plane1 at SFO)|} and {\small \verb|(Fly plane1 from SFO to JFK)|}. In the online case, the adversary removes actions from $\mathcal{O}$ during the execution of the plan (while the agent is interacting with the environment).

We call $H_{agent}$ the search heuristic used by the agent. Additionally, for an online planning system, we define $S_{current}$ as the current state of the agent at time $t_i$ and $S_{next}$ as the next state of the agent at time $t_{i+1}$. These values do not exist for an offline planner because the plan is computed beforehand.
We explore adversaries threat models including:
($a$) \textbf{Agent's Heuristic and Informed, Online} - This adversary knows the search heuristic used by the agent's planner (Euclidean, Manhattan, etc.).  Knowing the next state of the agent, $S_{next}$, means our adversary is {\it informed}. The adversary knows $S_{current}$, $S_{goal}$, $S_{next}$ and $H_{agent}$.
($b$) \textbf{Agent's Heuristic, Online} - This adversary knows the search heuristic used by the agent. However, $S_{next}$ is unknown. The adversary can only guess $S_{next}$ using the state with the best cost estimate given by $H_{agent}$. This guess can be incorrect if the agent's search algorithm is non-deterministic.
Here, the adversary knows $S_{current}$, $S_{goal}$, and $H_{agent}$.
($c$) \textbf{Black-Box, Online} - This adversary does not know anything concerning the agent's planner. Here, the adversary knows $S_{current}$ and $S_{goal}$.
($d$) \textbf{Agent's Heuristic, Offline} - This adversary knows the heuristic used by the planning system of the agent. The adversary knows $S_{init}$, $S_{goal}$, and $H_{agent}$.
($e$) \textbf{Black-Box, Offline} - This adversary does not know anything concerning the agent's planner, only $S_{init}$ and $S_{goal}$. The adversarial changes are specified at the beginning of the agent's computation. The agent needs to find a plan taking these changes into account.

\section{Approach} \label{Approach} 
\label{motivation_overview}

This section presents the window-heuristic to find adversarial changes. The window-heuristic outputs a set of $k$ grounded actions to remove the instance ($\mathcal{O}$) to increase the cost of the plan.

One could use the \textit{min-cut} algorithm to find a minimum cut of the state space. Here, we would partition the initial state and the goal state(s). The min-cut algorithm removes edges, which represent grounded actions in the state-space. When all the edges from the cut are removed by an adversary, there would be no path between $S_{init}$ and $S_{goal}$, meaning no plan actually exists and the adversary has succeeded. Unfortunately, the min-cut algorithm does not guarantee to cut less than $k$ edges, \ie less than the number of grounded actions that an adversary is able to prevent. The min-cut problem can be solved in polynomial time in the size of the input graph~\cite{MinCut}. However, in the general case, the state-space has an exponential number of nodes in the length of the task definition, thus making the min-cut algorithm inappropriate for finding adversarial examples.

One might alternately try all possible changes and keep the most adversarial one, \ie brute-force. More formally, given a task $T = (S_{init}, S_{goal}, \mathcal{O})$ we do the following: (1) Compute a plan for task $T$. (2) For every action $A_i$ in this plan, compute a solution of $(S_{init}, S_{goal}, \mathcal{O}\setminus\{A_i\})$. (3) Keep the most adversarial action $A_j$, \ie the one that maximally increases the cost of the initial plan.
This brute-force approach guarantees to find the best adversarial change if the plan computed in (1) is optimal. However, it requires that we run a planner $p+1$ times, where $p$ is the length of the initial plan. This is not practical because the search for a solution is generally computationally intensive~\cite{Complexity1, Complexity2}.

Even the most sophisticated planners can fail to find an existing solution. Moreover, the brute-force approach only outputs a single adversarial change. In order to output two adversarial changes, an adversary would have to run the planner once for every pair of actions in the initial plan, \ie{$p(p-1)+1$}. To output $k$ adversarial changes, the adversary would have to run their planner $O(p^k)$ times. Additionally, if the adversary does not find a solution to the initial problem during step (1), it is impossible to run step (2). Hence, brute-forcing is not practical in the general case: it is computationally intensive and assumes that an adversary has a planner as equally sophisticated as the agent.

We develop an approximation algorithm to output adversarial changes. We formally show that finding the $k$ adversarial changes is at least NP-Hard (Appendix~\ref{adversarialNP_HARD}). The window-heuristic enables us to bring planning instances of arbitrary size into a tractable scope.   The intuition is as follows: our strategy is to store known successful attacks from smaller (tractable) problems and project them onto larger (intractable) planning instances. Similar methods have been shown to enhance planning algorithms by reducing the number of nodes explored~\cite{patternDB}.
Here we model a large problem as a graph. Here, the goal for an adversary is to identify a small region of that graph (\ie a window) without being able to see the entire graph. To achieve this, the adversary will walk through the large graph to observe if the current region matches a previously observed region. Once a match is found, the adversary executes a known attack on the region.

Formally, we define a window as a connected sub-graph of the state-space, parameterized by $n$ (a state-space is a graph where the nodes are the states and the directed edges are the grounded actions). Hence, a window contains $n$ nodes (\ie states) and at least $n-1$ edges (\ie grounded actions) as it is connected. Figure~\ref{fig:window_viewAirCargo} is a visual representation of a window for the {\small \verb|air cargo transportation|} domain.
The formulation of our attack is divided into two parts: a generation phase and an execution phase.

\vspace{3pt}\noindent\textbf{Window Generation} - The adversary creates a table of windows, that are known to be adversarial (\ie when applied, the cost of the plan increases). To create the table, the adversary generates several \textit{simpler} planning instances (all from the same domain). The instances should be simple enough for the adversary to brute force them within the reduced state-space, \ie run step (1) and (2) from the previously explained brute-force algorithm (Section~\ref{Approach}). When the most adversarial action is found, we extract a window around it (Section~\ref{window_view} and~\ref{lookup_table}).

\vspace{3pt}\noindent\textbf{Window-Heuristic Execution} - Once the table is generated, the adversary is given a planning task and searches for adversarial changes with the window-heuristic (As described in Section~\ref{window_heuristic}). The adversary runs their own planning algorithm, computes a solution, and applies the window-heuristic. The adversary slides the windows from the table on the state-space. A match occurs when the windows are \textit{isomorphically equivalent} (further discussed in Section~\ref{lookup_table}), and thus, we output the associated grounded action (as shown in Fig.~\ref{fig:sliding_window}).

\subsection{The Window-View} \label{window_view}
An adversary links a window and an adversarial change. We say we \textit{apply} the window when we remove the grounded action associated with it from the set of grounded actions, $\mathcal{O}$. Intuitively, when we attack a planning instance, we apply a window when we see a matching one in the state-space of the arbitrary problem. Windows can take different shapes and sizes depending on the class domain.

\begin{figure}[!t]
\centerline{
\hfill
\subfloat[]{\includegraphics[width=0.25\linewidth]{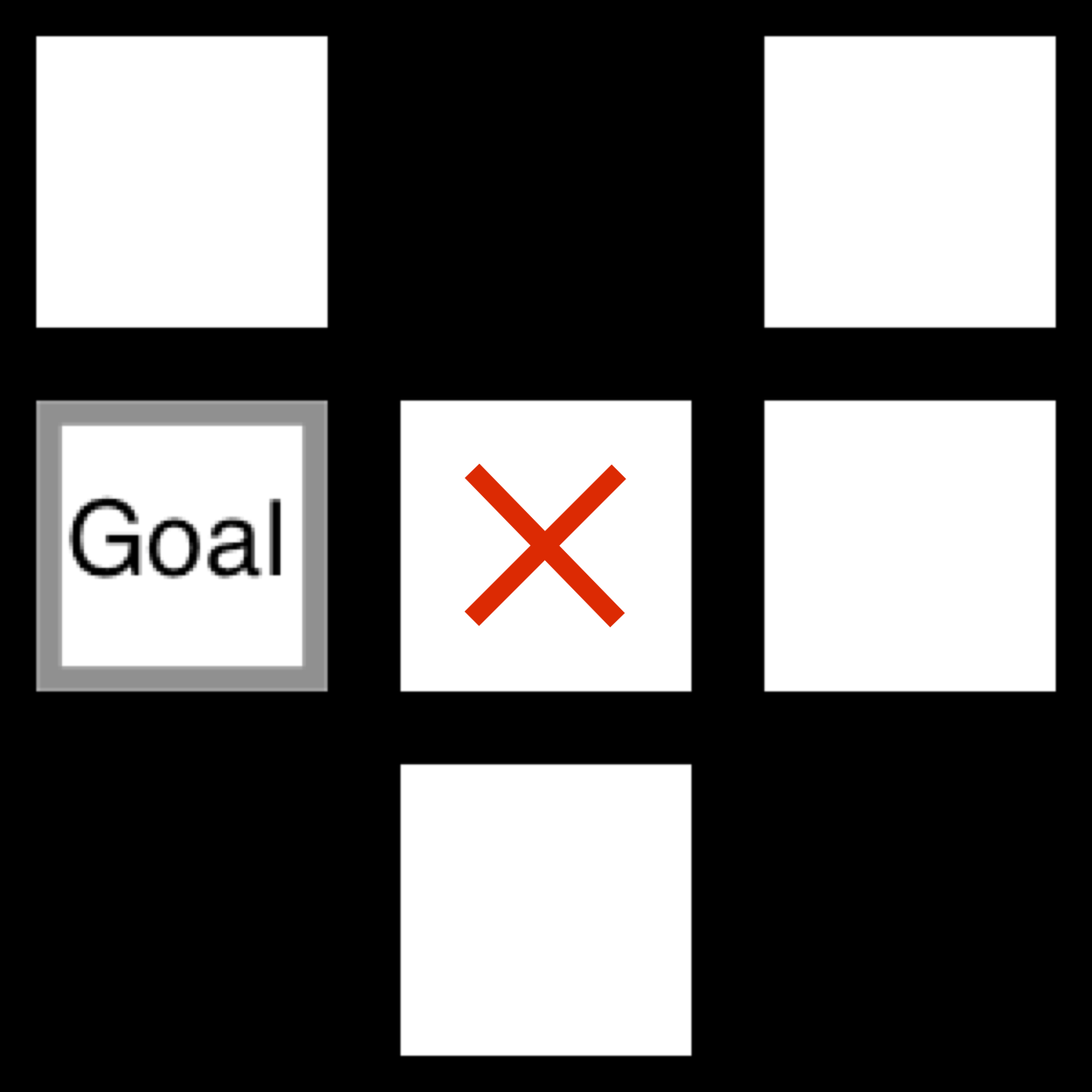}\label{fig:window_viewMazeA}}
\hfill
\subfloat[]{\includegraphics[width=0.25\linewidth]{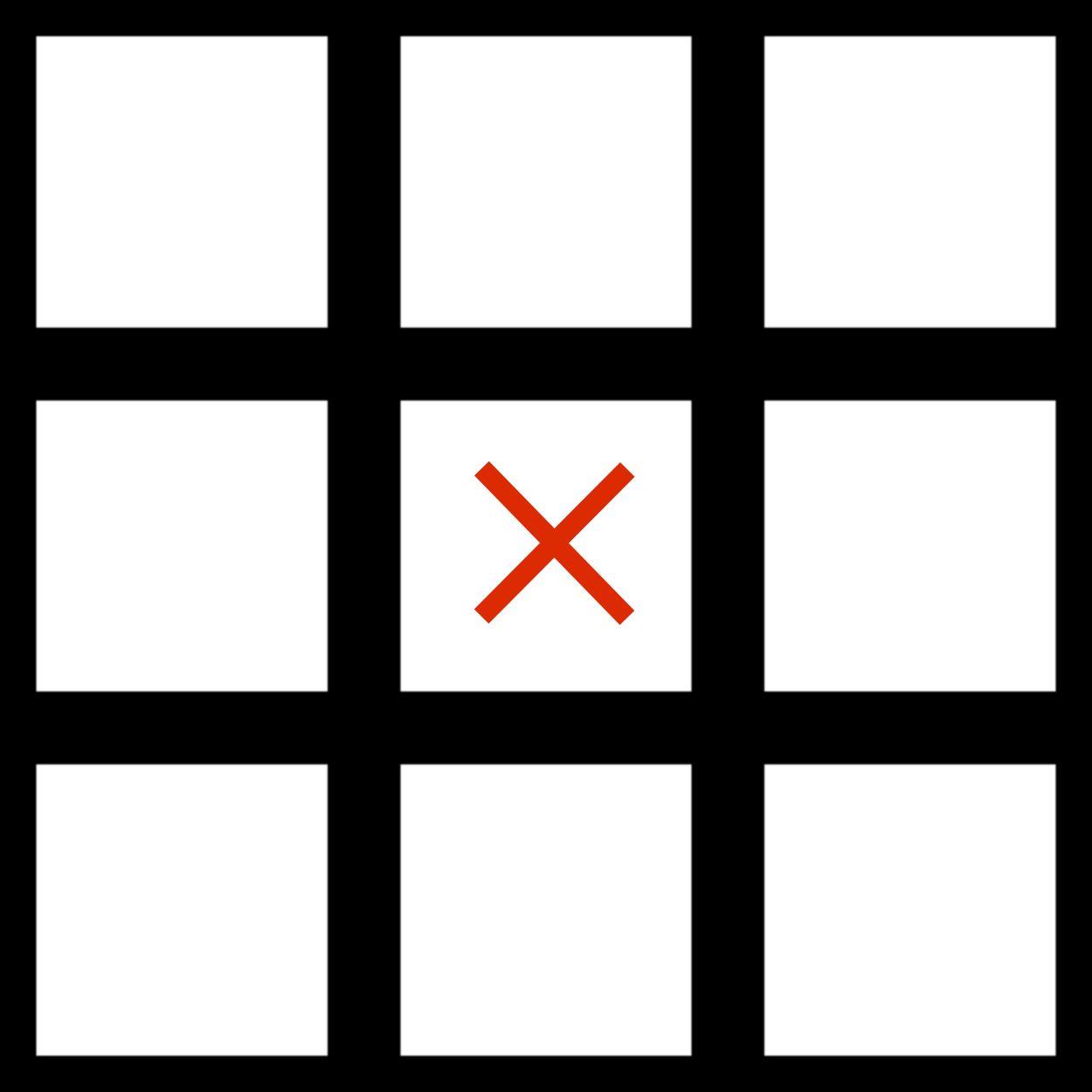}\label{fig:window_viewMazeB}}
\hfill
}
\centerline{\subfloat[]{\includegraphics[width=1\linewidth]{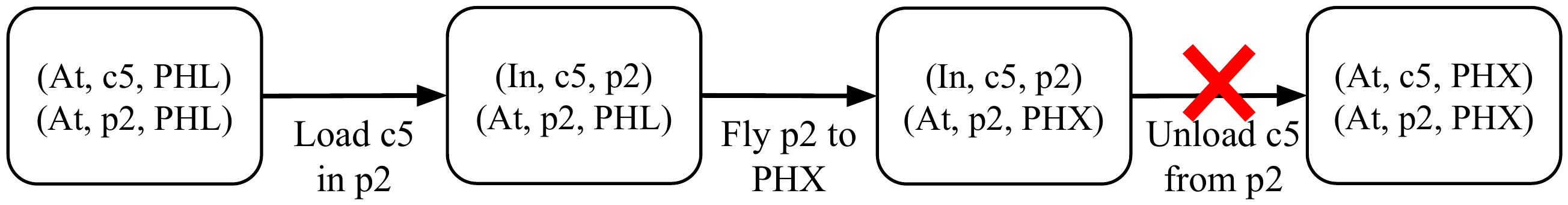}\label{fig:window_viewAirCargo}}}
\cprotect\caption{(a) An adversarial $3\times3$ window in path-finding. Placing a wall where the ``X'' is will prevent the agent from reaching the goal if it approaches from the bottom or the right of the window. As the only way to reach the goal is left, the overall path length is likely increased. (b) the (non-adversarial)  window is not likely to increase the cost of the plan. (c) An adversarial window ($n=4$, $4$ states, $3$ actions) from the {\small \verb|air cargo|} domain: removing the last action---{\small \verb|Unload c5 from p2 at PHX|} makes planning fail.}
\label{fig:window_view}
\end{figure}

For path-finding domains, we choose a window to be a local $n\times n$ node view of the agent's surroundings (Fig.~\ref{fig:window_viewMazeA}, \ref{fig:window_viewMazeB}). We define a \textit{wall} to be a node the agent cannot reach. The adversarial change associated with the $n\times n$ view is a wall at the center of the window. We apply the window when we see the same arrangement of walls in the environment. In this sense, to \textit{apply} a window means to add a wall at the center of the window.

For STRIPS-written problems, a window is a succession of $n$ states linked by $n-1$ actions. We apply a window when we see an equivalent succession of the $n-1$ first states in the state-space. The adversary applies a window by removing the last grounded action in that window from $\mathcal{O}$ (\ie the $n-1^{th}$). As shown in Figure~\ref{fig:window_viewAirCargo}, preventing (previously loaded) cargo from being unloaded is likely to be adversarial.

Window-size is chosen empirically. Ideally, it should be a function of the considered domain to maximize the success rate of the adversary. In the rest of this paper, we set $n=3$ for path-finding domains and $n=4$ for STRIPS domains.
Assuming a fixed number of entries in the table and a larger window size, finding an equivalent window is less likely to happen because it needs to be found over the entire sub-graph described by the window. On the other hand, with larger window sizes, the probability to perturb the plan when a match is found is increased. Indeed, the larger the window, the more alike the sub-graph on which we match the window has to be. The adversarial change within the window is more specialized and thus has a better chance to be adversarial.

\subsection{The Table of Advantageous Windows}
\label{lookup_table}

To create a table containing the most effective adversarial windows, the adversary generates several random \textit{simple} problems and extracts the most adversarial windows with an exhaustive search. Then, the adversary adds them to the table only if no other equivalent window (\ie isomorphic) is already in the table. Without an equivalence relation the adversary would end up with, potentially, an exponential number of entries in the table. The adversary can also limit the number of predicates in each node of a window using a \textit{normalization} process. When we extract a window for STRIPS instances, we capture $n$ states. Each of those states contains hundreds of predicates to describe the entire environment. However, because the environment does not change drastically within a window, many predicates remain unchanged across the $n$ states. Finally, while creating the table, we also compute  how frequently a window is adversarial. We can threshold the table to only keep the windows with the highest frequency. In doing so, we get a higher probability to increase the cost of the plan when we apply a window. Intuitively, adversarial windows that are frequently observed in smaller problems are more likely to increase the cost in larger problems.

Note that selecting random examples to generate windows is appropriate when the adversary has no knowledge of the instances expected at run time. In truth, table creation could be improved (perhaps vastly) by using examples of instances (or similar instances) likely to be encountered by the target at run time.  Indeed, in practice an intelligent adversary would collect known instances and ``train'' the window heuristic generation to find advantageous windows representative of those encountered by the victim system. We leave actively investigating other training approaches to future work.

\begin{figure}[!t]
\centering
\subfloat{\includegraphics[width=1\linewidth]{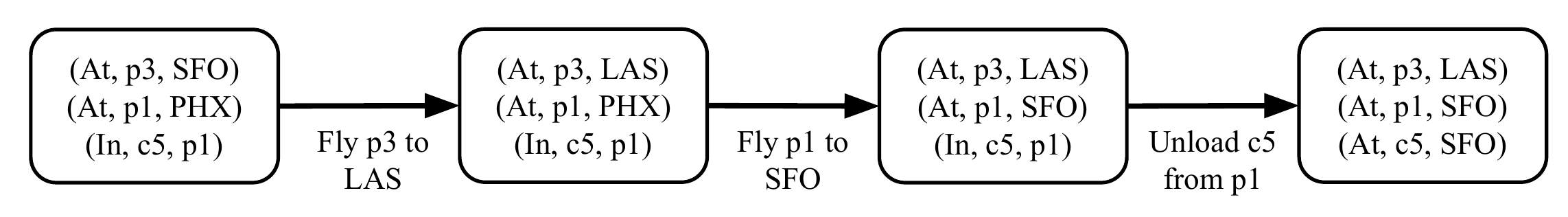}}
\\
\subfloat{\includegraphics[width=1\linewidth]{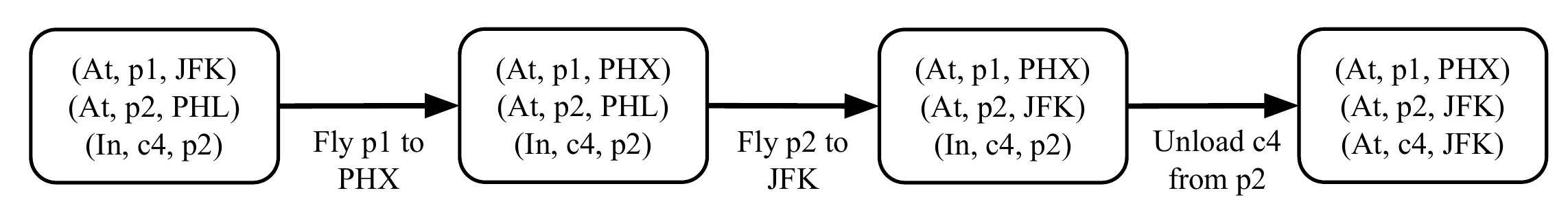}}
\cprotect\caption{Two equivalent windows for the {\small \verb|air cargo transportation|} domain. The actions are similar ({\small \verb|Fly|}, {\small \verb|Fly|}, {\small \verb|Unload|}) and there is a bijection between the objects. The bijection is the following: $f(p1) = p3$, $f(JFK)=SFO$, $f(PHX)=LAS$, $f(p2)=p1$, $f(PHL)=PHX$, and $f(c4)=c5$.}
\label{window_equivalence}
\end{figure}

\vspace{3pt}\noindent\textbf{Graph Isomorphism} - The predicates describing a state are grounded, meaning they do not contain any free variable. A state would contain the predicate {\small \verb|(At, p1, JFK)|} instead of {\small \verb|(At, plane, airport)|}. This notation is dependent on how we choose to name the airports and the planes. We need an equivalence relation that does not rely on the objects' names. Consider the following predicates {\small \verb|(At, p1, JFK)|} and {\small \verb|(At, plane1, airport1)|}. Renaming {\small \verb|p1|} to {\small \verb|plane1|} and {\small \verb|JFK|} to {\small \verb|airport1|} gives us the equivalence. For STRIPS-written tasks, we say that window $w_1$ is isomorphically equivalent to window $w_2$ if (1) there is a bijection $f$ between the name of the objects such that $w_1 = f(w_2)$, {\small \verb|plane1|} $=f(${\small \verb|p1|}$)$ and (2) $w_1$ and $w_2$ share the same non-grounded operators (Fig.~\ref{window_equivalence}). For path-finding, we say that two windows are equivalent if there is a rotation $r$ such that $w_1 = r(w_2)$.

\begin{figure}[!t]
\centering
\subfloat{\includegraphics[width=1\linewidth]{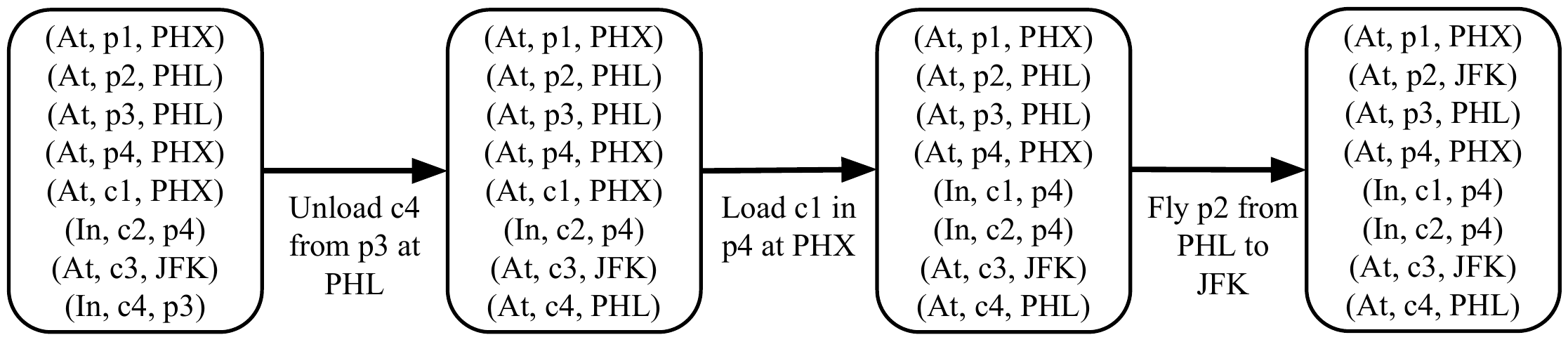}}\\
\subfloat{\includegraphics[width=1\linewidth]{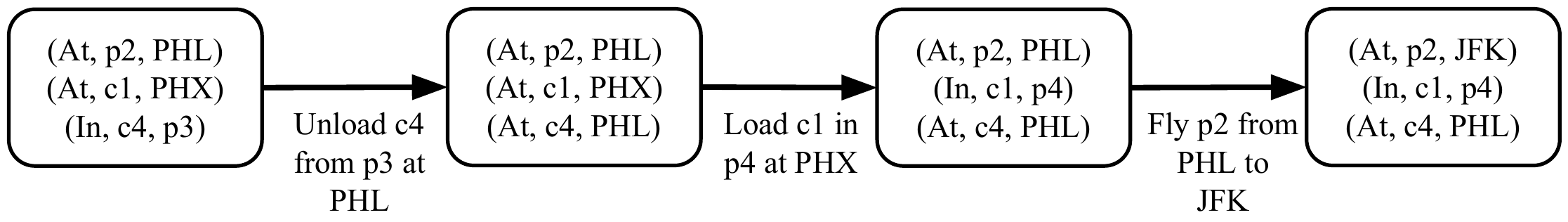}}
\cprotect\caption{The window before normalization (top) and after (bottom). All the predicates that weren't modified by the three actions were removed, e.g., {\small \verb|(At, p1, PHX)|} is removed.}
\label{window_normalization}
\end{figure}

\vspace{3pt}\noindent\textbf{Normalizing} - Normalizing is a process we only perform for STRIPS windows because path-finding states are not defined with predicates. We normalize to remove constant predicates in an extracted window. Consider a window ($n=4$) containing four states and three grounded actions extracted from the {\small \verb|air cargo transportation|} domain. The planning task can contain an arbitrary number of {\small \verb|plane|} objects. Each plane would need a predicate to indicate its position: {\small \verb|(At, plane, airport)|}. However, if a plane does not move, the predicates concerning its position do not change and are repeated across the four states. Thus, they are not relevant to the evolution of the environment within the four states. In Figure~\ref{window_normalization}, none of the three actions affect plane {\small \verb|p1|}, so we remove all the predicates concerning {\small \verb|p1|} in the window. The only predicates that are going to change between the four states are the ones modified (\ie added, deleted) by the three grounded actions. More formally, we call the four states in the window $S_{\llbracket1;4\rrbracket}$. Each $S_{i}$ is a set of predicates. We introduce $\Delta = \bigcap_{i=\llbracket1;4\rrbracket} S_{i}$ and we define $\forall i \in \llbracket1;4\rrbracket, \hat{S_{i}} = S_{i} \setminus \Delta$. The $\hat{S_{i}}$ are the new normalized states where we removed the redundant predicates.

\vspace{3pt}\noindent\textbf{Thresholding} - With the normalizing process and isomorphisms, we drastically decrease the size and the number of entries needed in the table. We introduce the threshold in order to only keep the windows that appeared the most in the table. Indeed, a window might be adversarial for 50\% of the problems, and another one might be adversarial for 1\% of the problems. Still, both of those windows appear in the table with the same importance. To address this, once the table is filled, we empirically select only the most frequently observed windows. If a window is adversarial once over a thousand tasks, there is no need to record it---it is too specific.

The formal process to create the table is the following (Algorithm~\ref{Alg:create_table}). (1) Generate several random \textit{simple} tasks from of the domain considered. (2) Find the most malicious windows using a brute-force approach for all of those  (line~\ref{create_table:BruteStart} to \ref{create_table:BruteEnd}). (3) Extract a window around the adversarial change with the highest cost increase. (4) Normalize the window if needed and add it to the table (line~\ref{create_table:AddStart} to \ref{create_table:AddEnd}). Finally, (5) threshold the table to only keep the most adversarial windows (line~\ref{create_table:thresh}).

\begin{algorithm}[!t]
\small
\For{$i\leftarrow 1$ \KwTo $N$}{
    Generate a random problem $P$\;
    Solve problem $P$ with a planner\;\label{create_table:Solve}
    Let $C$ be its cost and $PL$ the plan\;\label{create_table:BruteStart}
    Let $bestAdvC$ be the best adversarial cost\;
    Let $bestAdvAct$ be the best adversarial action\;
    \ForEach{action $A$ in the plan $PL$}{
        Solve problem $P$ without the action $A$ allowed\;
        Update $bestAdvC$ and $bestAdvAct$\;
    }\label{create_table:BruteEnd}
    \If{$bestAdvC > C$}{
        Take a window $W$ around $bestAdvAct$\; \label{create_table:AddStart}
        Normalize the window $W$\;
        \eIf{$W$ has an equivalent in the table}{
            Add $1$ to the number of occurrences of $W$\;
        }{
            Add $W$ to the table\;
        }\label{create_table:AddEnd}
    }
}
Threshold the table\;\label{create_table:thresh}
\textbf{return} the table\;
\caption{Constructing a table of advantageous adversarial windows using N random \textit{simple} problems. For each of those problem we extract the most adversarial window and add it to the table.}
\label{Alg:create_table}
\end{algorithm}

\subsection{The Window-heuristic} \label{window_heuristic}

We now explain how to use the table to run the window-heuristic and output adversarial changes. An adversary and the agent are given an arbitrary size problem $(S_{init}, S_{goal}, \mathcal{O})$. The goal for the agent is to find an optimal plan. The goal for an adversary is to output the best set of grounded actions to perturb the plan. We distinguish the offline case (STRIPS-written problems) from the online case (path-finding). Those two cases differ in the attack scenario: in the offline case, the goal for an adversary is to output $k$ adversarial changes before the agent starts planning. For the online case, an adversary applies these changes directly and perturbs the agent's environment while the agent is planning. That means the agent can react (find another plan) to an adversary's changes.

\begin{figure}[!t]
\centering
\includegraphics[width=0.9\textwidth]{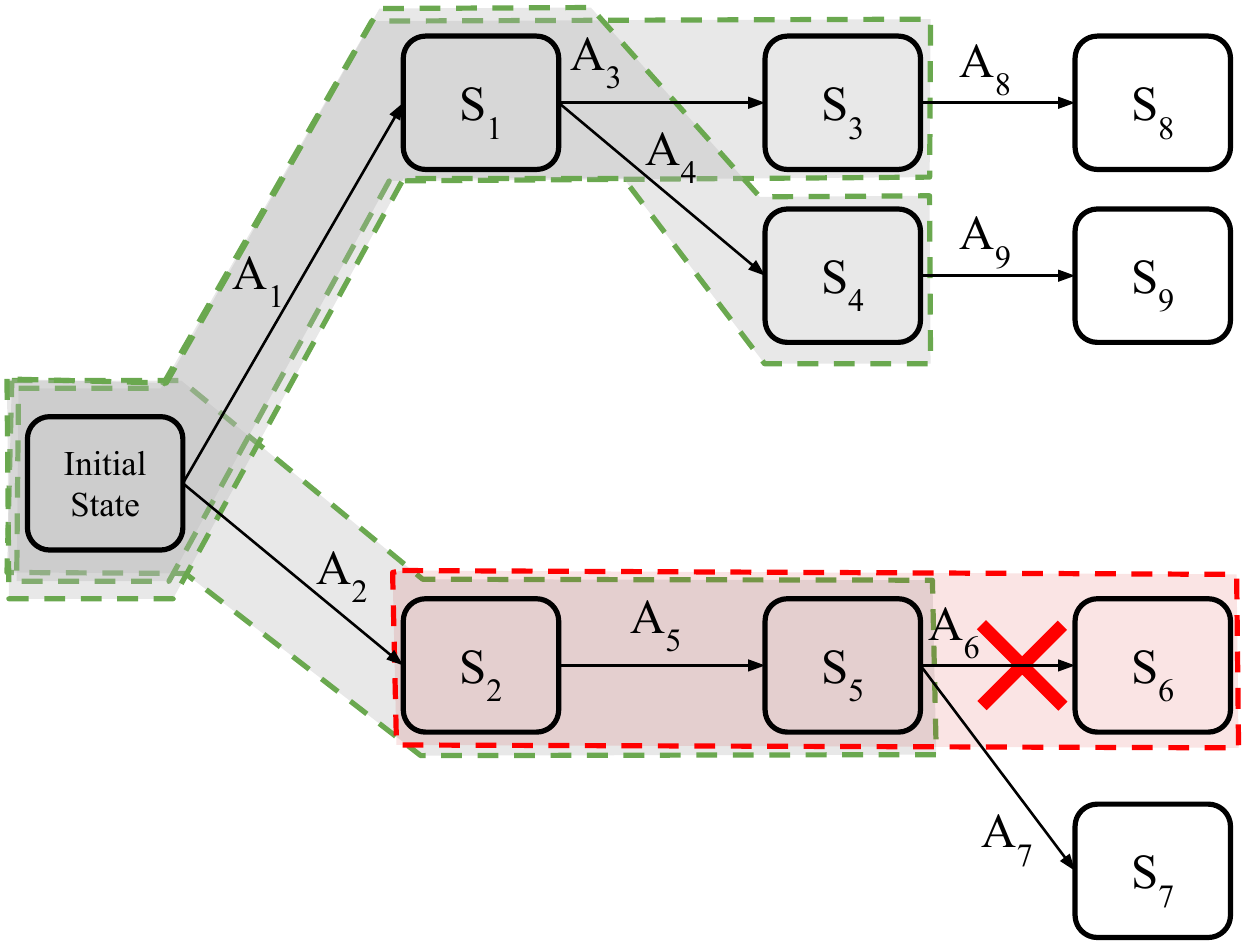}
\cprotect\caption{An adversary expands the state space starting with the initial state and discovers $S_1$ and $S_2$. Assume $H_{adv}(S_1) < H_{adv}(S_2)$, then $S_1$ is the next state to expand. After discovering $S_1$'s successors, an adversary checks if the windows $S_{init}-A_1-S_1-A_3-S_3$ and $S_{init}-A_1-S_1-A_4-S_4$ have a match in the table. Assuming no match is found, it continues until a match is discovered at $S_2-A_5-S_5-A_6-S_6$.}
\label{fig:sliding_window}
\end{figure}

\vspace{3pt}\noindent\textbf{Offline Window-heuristic} - This procedure is detailed in Algorithm~\ref{Alg:Heuristic_basic_scheme_offline}. An adversary explores the state-space by running a separate planner, to find the solution to the input task (Recall, the adversary knows at least the initial and goal states). In doing so, an adversary expects to expand the state-space in the same direction(s) as the agent. If an adversary successfully predicts the state-space expansion, applying windows hindering that expansion will be adversarial for the agent. The search starts from $S_{init}$ and stops either when an adversary's planner returns or when $k$ adversarial changes have been found. Essentially, when an adversary recognizes a window from the table during the state-space expansion, the adversarial change associated is applied. During the search, we extract a window around the next state to be expanded by the adversary's planner: $S_{next}^{adv}$ (line~\ref{Heuristic_basic_scheme_offline:line:Take}). We check if this window has a match in the table. If it does, we apply the adversarial change to the planning task (line~\ref{Heuristic_basic_scheme_offline:line:Remove}). In Figure~\ref{fig:sliding_window}, we show how the adversarial change associated with the window recognized is removed from the plan.

\begin{algorithm}[!t]
\small
\KwIn{Number of adversarial changes allowed $k$;}
\nonl\myinput{The initial state $S_{init}$;}
\nonl\myinput{The goal state characterization $S_{goal}$;}
\nonl\myinput{The set of operators $\mathcal{O}$;}
\nonl\myinput{The advantageous windows table $Wtable$;}
\KwOut{A set of grounded actions $S$;}
\BlankLine

$S_{expanded}$ = $S_{init}$\;
\While{$S_{expanded} \neq S_{goal}$ \textbf{and} $|S| < k$}{
    Find $S_{next}^{adv}$ with search algorithm\;
    Take a window $W$ around $S_{next}^{adv}$\;\label{Heuristic_basic_scheme_offline:line:Take}
    Normalize $W$\;
    Search for an equivalent of $W$ in $Wtable$\;
    \If{an equivalent has been found}
    {
        Extract the adversarial change from $W$\;
        Add it to the the set $S$\;
        Remove the action from $\mathcal{O}$\;\label{Heuristic_basic_scheme_offline:line:Remove}
        Find a new $S_{next}^{adv}$ with search algorithm\;
    }
    $S_{expanded}$ = $S_{next}^{adv}$
}
\textbf{return} $S$\;

\caption{Offline window-heuristic: Given an offline planning instance it outputs a set of adversarial changes.}
\label{Alg:Heuristic_basic_scheme_offline}
\end{algorithm}

\vspace{3pt}\noindent\textbf{Online Window-heuristic} - For the online case, we do not need to pre-determine a set of adversarial changes. The adversarial changes are applied at run-time and the agent needs to adapt. The window-heuristic for an online planning task runs about the same way except an adversary does not run a separate planner. Indeed, there is no need to predict the agent's entire state-space expansion because we assume an adversary knows $S_{current}$ at all times. $S_{current}$ is the position of the agent and is updated as the agent moves (i.e. executes actions in the environment). We only need to predict the next state to be visited by the agent $S_{next}$ where $S_{next}$ is defined as:
\begin{equation}
    S_{next}=\argmin_{S \in successor(S_{current})}[H_{agent}(S)]
\end{equation}

Each time the agent moves and updates $S_{current}$, an adversary computes an estimation of $S_{next}$ (line~\ref{Heuristic_basic_scheme_online:line:Predict}). In order to estimate $S_{next}$, an adversary uses the search heuristic $H_{adv}$.
The state $S_{next}^{adv}$ predicted by an adversary is $S_{current}$'s neighbor with the lowest $H_{adv}$ image:
\begin{equation}
    S_{next}^{adv}=\argmin_{S \in successor(S_{current})}[H_{adv}(S)]
\end{equation}
 An estimation of $S_{next}$ is calculated, the adversary can apply the same mechanism as the offline case. If the adversary recognizes a window around $S_{next}^{adv}$, we apply the adversarial change  (line~\ref{Heuristic_basic_scheme_online:line:Apply}).

\begin{algorithm}[!t]
\small
\KwIn{Number of adversarial changes allowed $k$;}
\nonl\myinput{The current state of the agent $S_{current}$;}
\nonl\myinput{The goal state characterization $S_{goal}$;}
\nonl\myinput{The set of operators $\mathcal{O}$;}
\nonl\myinput{The lookup table $Wtable$;}
\BlankLine

$A_c \gets 0$\;
\While{$S_{current} \neq S_{goal}$ \textbf{and} $A_c < k$}{
    Estimate $S_{next}$ using $H_{adv}$, $S_{current}$ and $S_{goal}$\; \label{Heuristic_basic_scheme_online:line:Predict}
    Take a window $W$ around $S_{next}^{adv}$\;
    Normalize $W$\;
    Search for an equivalent of $W$ in $Wtable$\;
    \If{an equivalent has been found}
    {
        Extract the adversarial change from $W$\;
        Apply the adversarial change\;\label{Heuristic_basic_scheme_online:line:Apply}
        $A_c \gets A_c+1$\;
        \tcp{\small The agent has to re-plan.}
    }
    Wait for $S_{current}$ to change\;
    \tcp{\small The agent has re-planned and moved.}
}
\caption{Online window-heuristic: Given an online planning instance it perturbs the agent's plan at run-time.}
\label{Alg:Heuristic_basic_scheme_online}
\end{algorithm}

\section{Evaluation} \label{Evaluation}

In this section, we evaluate our approach in several planners and domains for both online and offline attacks.  Table~\ref{table:Evaluation_scenarios} summarizes the experiments and results. We begin by exploring D*Lite. We ask:

\begin{table*}[!t]
\begin{center}
\begin{threeparttable}
{\small
 \begin{tabular}{|c|c|c|c|c|}
 \hline
 Planning type & Domain & Agent's planner & Threat model & Success Rate \\ [0.5ex]
 \hline\hline
 Online & Maze & D* Lite & Agent's heuristic and Informed & 66.86\% - 82.65\% \, \\
 Online & Maze & D* Lite & Agent's heuristic & 57.53\% - 79.73\%\, \\
 Online & Maze & D* Lite & Black-box & 60.41\% - 82.75\%\,\\
 Offline & Barman & Fast Downward & Black-box & 85.71\% \, \\
 Offline & Floortile & Fast Downward & Black-box & 95.00\% \, \\
 Offline & Hiking & Fast Downward & Black-box & 75.00\%\, \\
 Offline & Tetris & Fast Downward & Black-box & 70.59\% \, \\
 Offline & Airport & Fast Downward & Black-box & 100.00\% \,\\
 Offline & Openstacks & Fast Downward & Black-box & 100.00\% \,\\
 Offline & Data-network & Fast Downward & Black-box & 41.67\%\,\\

\hline
\end{tabular}
\caption{All threat models and domains considered for the evaluation. Ranges in the success rates signify the 1 wall to 2 wall success rate while concrete values are when an adversary has up to 4 grounded actions.}
\label{table:Evaluation_scenarios}
}
\end{threeparttable}
\end{center}
\end{table*}

\begin{anonsuppress}

\begin{enumerate}
    \item Given an arbitrary planning task, how often can an adversary increase the cost? (\ie the success rate)
    \item Given a set of instances what is the average cost increase an adversary can expect to generate?
    \item What is the minimum knowledge an adversary needs in order to perturb a planning algorithm?
\end{enumerate}

First, note that not all planning tasks are sensitive to adversarial changes. There may be multiple disjoint plans to reach a goal state. Still, as we found, an adversary---depending on their knowledge---can expect a success rate from 70\% to 95\% with up to 4 adversarial changes. This value highly depends on the domain and the generation phase parameters (threshold, number of tasks generated, nature of tasks generated, window size...). On average, with up to 4 adversarial changes, an adversary will produce a plan from 1 to 10 steps longer (when the task is still solvable). For the black-box online attack (defined in Section~\ref{Adversarial_capabilities}) for path-finding tasks in mazes, we achieve a success rate of 56.52\% with one wall and 78.98\% with two walls (Section~\ref{Eval_Mazes}).

\end{anonsuppress}

\subsection{D*Lite}\label{Eval_Mazes}

\begin{figure*}[!htb]
\captionsetup[subfloat]{farskip=0pt,captionskip=0pt}
\hspace{-0.3in}
\subfloat[]{\includegraphics[width=.33\textwidth]{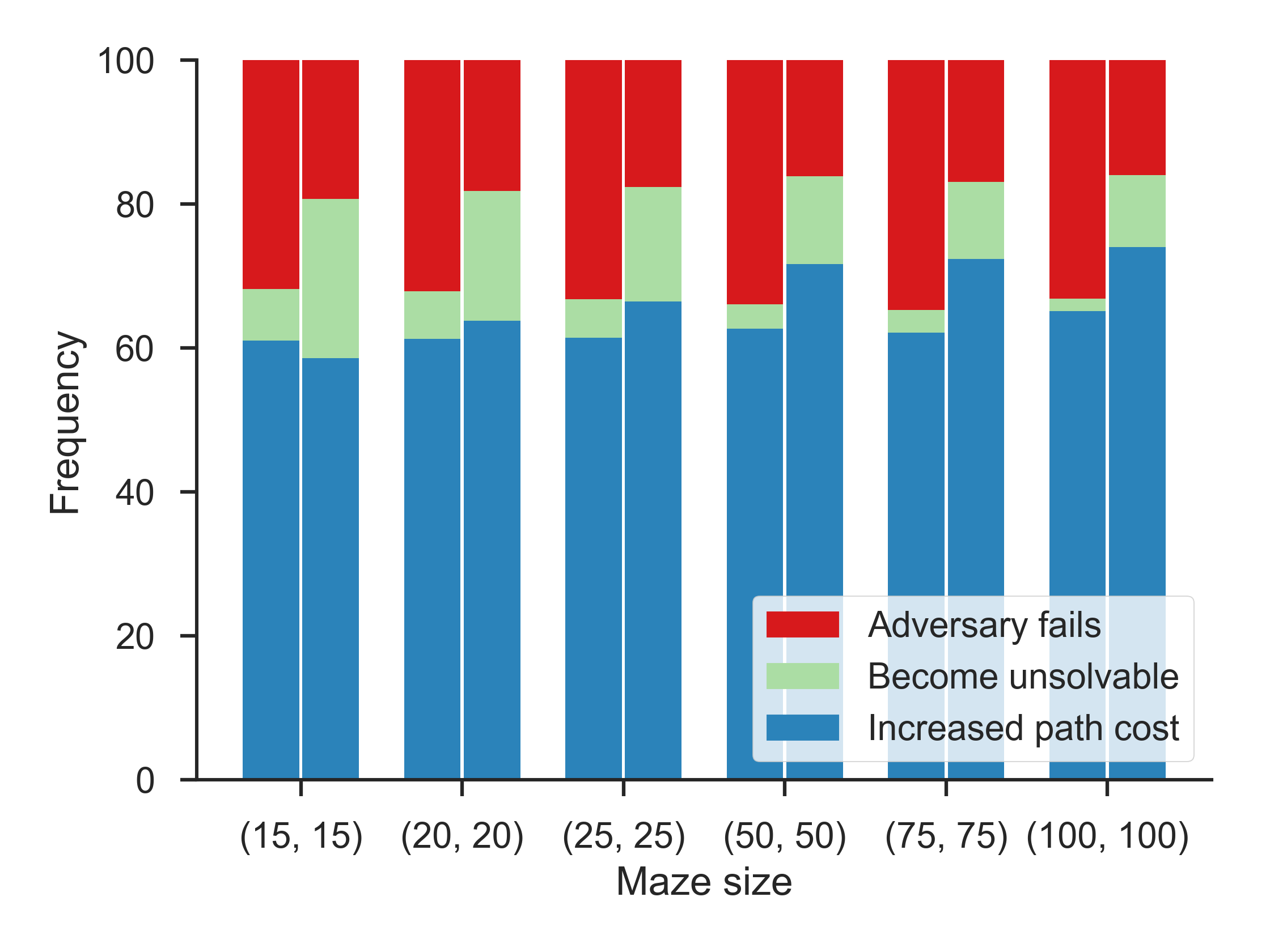}\label{Maze_Evaluation_Informed_Bar}}
\subfloat[]{\includegraphics[width=.33\textwidth]{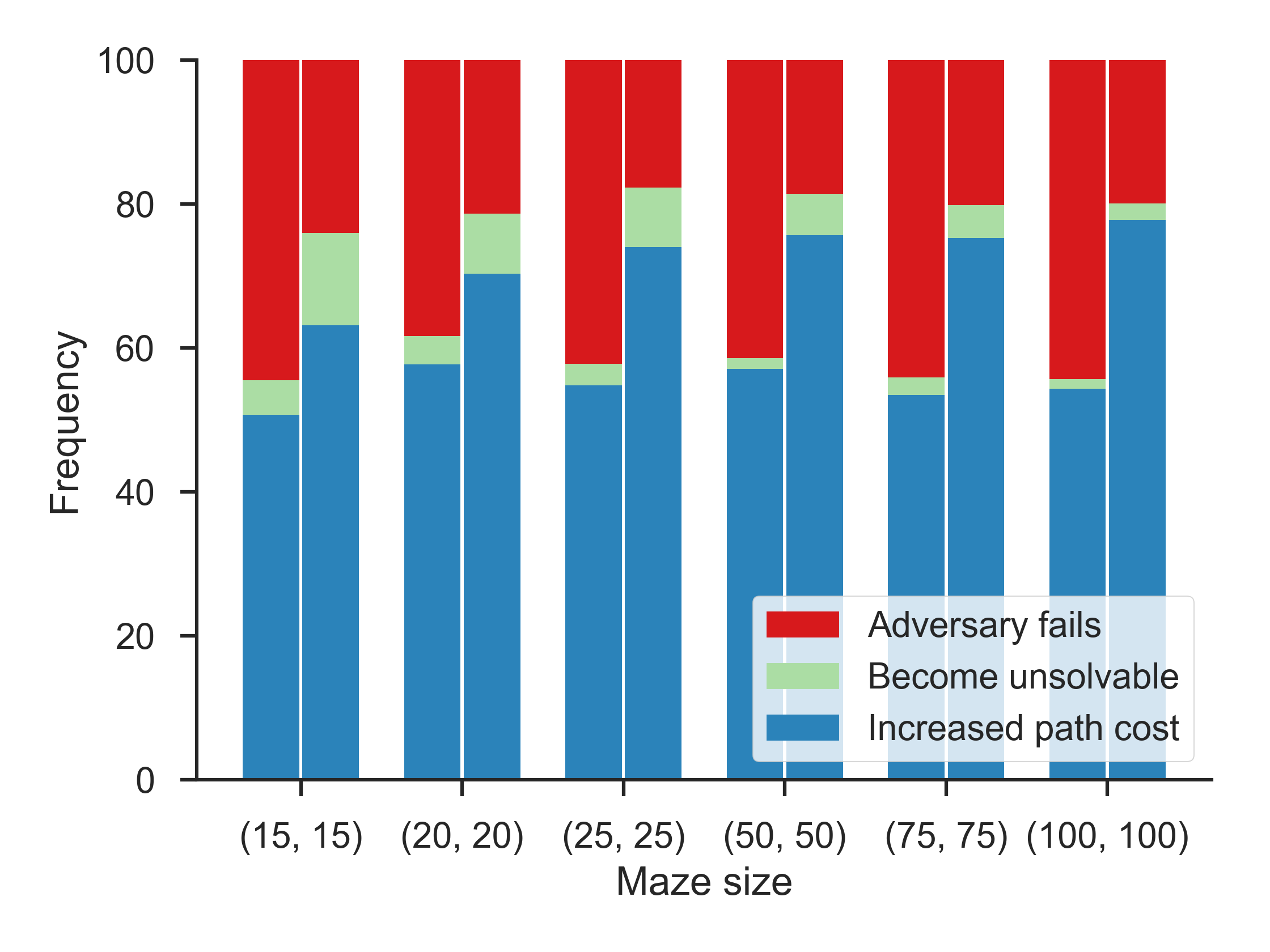}\label{Maze_Evaluation_Heuristic_Bar}}
\subfloat[]{\includegraphics[width=.33\textwidth]{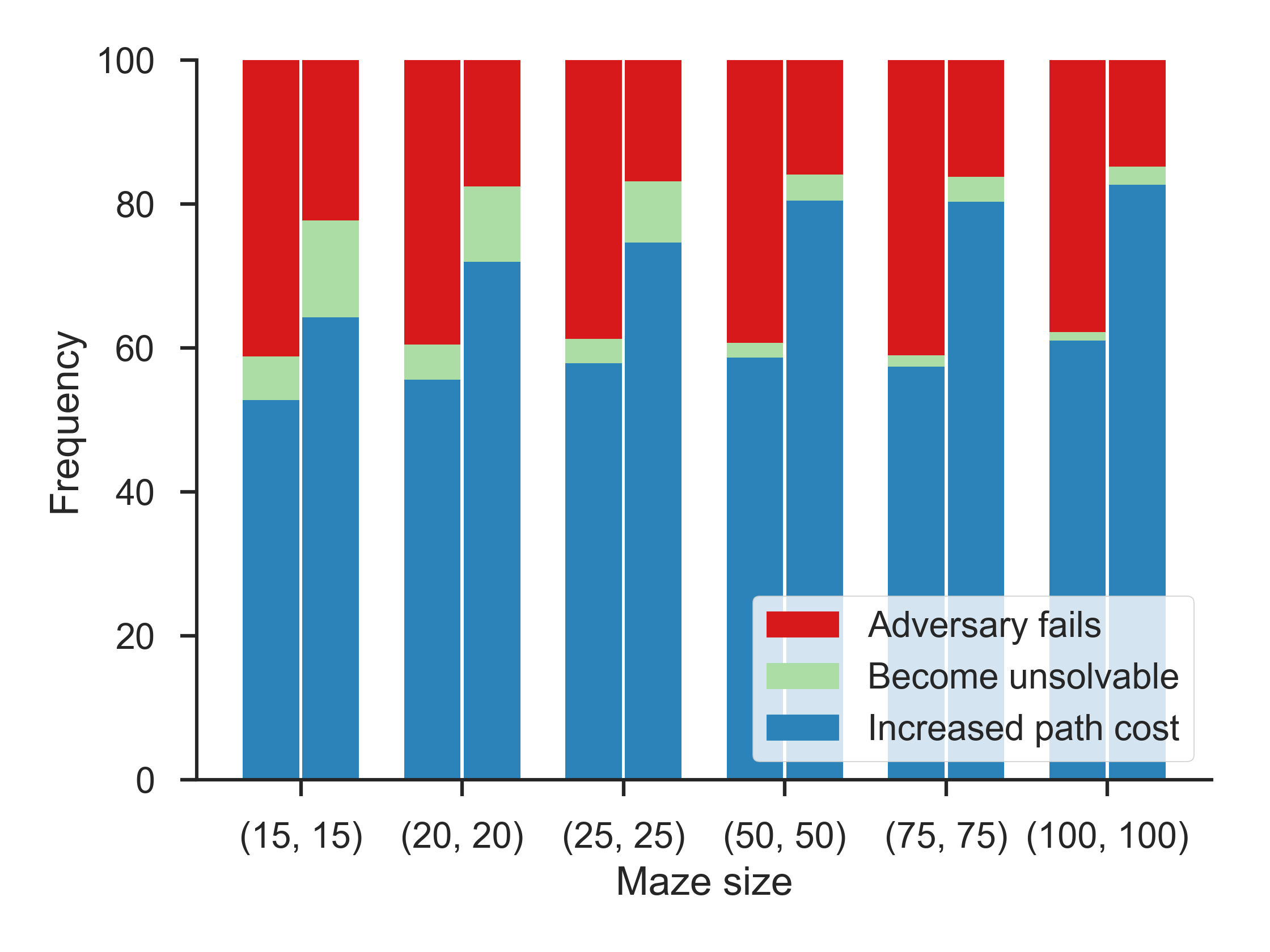}\label{Maze_Evaluation_Blackbox_Bar}}

\caption{The three bar graphs show the success of an adversary placing walls in mazes. The threat models are the following (a) Agent's heuristic and Informed (b) Agent's heuristic (c) Black-box. The bars on the left (resp. right) describe the success rate of an adversary capable of placing one wall (resp. two walls).}
\label{Maze_Evaluation}
\end{figure*}

We evaluate the success rate of the window-heuristic for an agent using the path-finding algorithm D*Lite\cite{D*Lite} in a maze. Here are the details of the experimental setup:
\begin{itemize}
    \setlength{\itemsep}{1pt}
    \setlength{\leftmargini}{1.3em}
    \item The agent uses D*Lite combined with the Euclidean distance to guide the search for a solution: $h_{Euclidean}(s) = \sqrt{(s_x-goal_x)^2+(s_y-goal_y)^2}$. When the agent's heuristic is unknown (Black-box scenario), an adversary uses the Manhattan distance: $h_{Manhattan}(s) = |s_x-goal_x|+|s_y-goal_y|$.
    \item There is only one goal state in the maze ($S_{goal}$). The goal for the agent is to reach it from the initial state ($S_{init}$).
    \item The agent is not allowed to move in diagonals, the only actions authorized are to move up, down, left, or right.
\end{itemize}

The goal of an adversary is to select the optimal location in a maze to place a wall in order to maximally increase the length of the path. The first step is to create the table of windows explained in Section~\ref{lookup_table}. Recall that the creation of the table is an offline process. To do so, we generate 500 random mazes (size $15\times15$ - wall frequency $0.25$) and we brute-force the best adversarial wall. For each tile on the initial path, we try to place a wall and we compute the path cost of the modified maze. We extract a $3\times3$ window around the most adversarial wall and we add it to the table.
Appendix~\ref{appdx:calib}, shows the empirical settings for the generation phase that gave the best experimental results (success rate, average path cost increase).

Once the table of advantageous adversarial windows is created, we can run the window-heuristic. The agent is given a maze instance and tries to reach a goal position. The adversary places walls as the agent moves increasing the number of steps required for the agent. We distinguish the three different online scenarios.

\vspace{3pt}\noindent\textbf{Agent's Heuristic and Informed, Online} - We start with a powerful adversary knowing $H_{agent}$ and $S_{next}$.  Figure~\ref{Maze_Evaluation_Informed_Bar} shows an increase the cost of the plan by adding $1$ wall in 66.86\% of all cases. Note that not all mazes can have their cost perturbed; sometimes one wall is not enough if two or more disjoint but equally optimal paths exist. We reach a success rate of 82.65\% when the adversary has the opportunity to place two walls in the way of the agent. On average, we increase the path by $2.48$ steps when the adversary adds $1$ wall and by $4.62$ steps when allowed to place up to $2$ walls. 

\looseness=-1
\vspace{3pt}\noindent\textbf{Agent's Heuristic, Online} - This scenario is harder for the adversary and more realistic. The adversary only has access to the heuristic the agent is using, its current position, and the goal state. Since we do not know which way the agent is going in that case, the adversary has to estimate the next move of the agent using $H_{agent}$. With one wall we modify on average 57.53\% (Fig.~\ref{Maze_Evaluation_Heuristic_Bar}) of the plans and 79.73\% with two walls (all sizes of mazes considered). As we can expect the efficiency drops, but by 5-10\% which is still a notable success rate because statistically, the adversary still manages to increase the cost. We increase it by $2.27$ steps with $1$ wall and by $4.44$ with $2$ walls.

\vspace{3pt}\noindent\textbf{Black-Box, Online} - The adversary does not have access to the heuristic the agent is using. The adversary has to create the table and predict the $S_{next}$ using a different heuristic. We choose $H_{adv}=h_{Manhattan}\neq H_{agent}$. This will answer the question about the transferability (between $H_{adv}$ and $H_{agent}$) of the attack. The agent however solves the maze using the Euclidean distance. With $1$ wall the adversary achieves a success rate of 60.41\% and increase the cost on average by $2.53$ steps (Fig.~\ref{Maze_Evaluation_Blackbox_Bar}). With $2$ walls we have a high success rate of 82.75\% and a cost increase of $5.48$ steps.

\vspace{3pt}\noindent\underline{\textbf{Takeaways}} - To summarize, we found that an adversary equipped with the window-heuristic is able to successfully increase the plan cost. The adversary needs little knowledge to perturb these kinds of path-finding problems. The difference between the success rate of the black-box and the informed scenario is small. Intuitively, the estimation of $S_{next}$ in the black-box scenario is often correct, given the limited number of actions the agent can take. Also, the window-heuristic scales well with the problem size and the success rate stays constant with different maze sizes.

All of the experiments here considered an online setting. We now move to an evaluation of the offline setting with an adversary running the heuristic before the agent starts to plan.

\subsection{Fast Downward}
\label{sec:fdown}
We show the efficiency of the window-heuristic on diverse tasks solved by the Fast Downward (FD) planner. The planner solves STRIPS instances with a large variety of settings. The user can specify the search strategy (\eg A*, Greedy search, hill-climbing...) and the search heuristic---called evaluator (\eg FF, CEA, and Landmark-count). Fast Downward also translates the STRIPS instance into other data structures to enhance the search for a solution. The Fast Downward planner solves a planning instance in three phases: translation, knowledge compilation, and search. During the translation phase, the FD planner performs grounding of predicates and operators~\cite{FastDownward}, basically creating the set of grounded operators, $\mathcal{O}$. Concretely, our algorithm removes $k$ grounded actions from $\mathcal{O}$ at translation phase. In practice, an adversary would have to prevent those actions.

The planner the adversary can run is a simple forward search planner running breadth-first search (BFS) or A* coupled with a single heuristic (Additive Cost ~\cite{Bonet01planningas}, etc.). This self-made planner run by the adversary is less sophisticated compared to FD. We evaluate an attack in which the adversary does not know how the FD planner works; only the domain specifications, $S_{init}$ and $S_{goal}$ are known. This attack is considered to be \textit{black-box offline}.

We evaluate the window-heuristic a in the following steps: ($a$) We generate tables as described in  Table~\ref{table:FD_generation} ($b$) and run the FD planner to output a plan without the adversary interfering. The agent runs FD with the FF heuristic~\cite{FFHeuristic} and context-enhanced additive heuristic~\cite{CEAHeuristic} with lazy best-first search and preferred operators.\footnote{This was one of the best configurations available according to the benchmarks run by Helmert in 2006~\cite{FastDownward}.} ($c$) We run the window-heuristic with the adversary's planner to output $k$ adversarial change(s), i.e., the grounded operator(s) that will be removed from $\mathcal{O}$ ($d$) We run FD again (same settings) with the limited set of operators. ($e$) Finally, we compare the cost of the plan with and without an adversary.

\begin{table}[!t]
\begin{center}
 \begin{tabular}{|c|c|c|c|c|c|}
 \hline
 Domain & Problem\, & Threshold & Window\, & Algorithm\, & $H_{adv}$\\ [0.5ex]
 \hline\hline
Barman & 200 & 10 & 5 & A* & Additive cost\\
Floortile & 200 & 1 & 35 & A* & Additive cost\\
Hiking & 100 & 3 & 11 & BFS & $\emptyset$ \\
Tetris & 200 & 1 & 24 & A* & $h_{CmonPred}$\\
Airport & 15 & 5 & 8 & A* & Additive cost\\
Openstacks & 10 & 0 & 34 & A* & Additive cost\\
Data-network & 15 & 0 & 19 & A* & $h_{CmonPred}$\\
\hline
\end{tabular}
\caption{The generation process' settings across domains. The problem corresponds to the number of problems generated to fill the table, the window is the number of windows kept in the table, and the algorithm is the algorithm used by the adversary. $h_{CmonPred}(s)$ is defined as  $|goal_{pred}|-|goal_{pred} \cap s_{pred}|$.}
\label{table:FD_generation}
\end{center}
\end{table}

We benchmark tasks and domains including several FD-based planners that performed the best during the 2014 IPC competition as well as several real-world applications:

\begin{itemize}

    \item The {\small \verb|airport|} domain tasks aim to control the ground traffic at an airport. Airplanes must reach their destination gate. There is outbound and inbound traffic; the former are airplanes that must take off, the latter are airplanes that have just landed and have to park~\cite{IPC4_Domains}. The instances we ran were based on the Munich Airport\footnote{Hatzack developed a realistic simulation tool, which he supplied to the IPC organizers to generate the domain instances. The simulator included Frankfurt, Zurich, and Munich airports. Frankfurt and Zurich proved too large for IPC purposes~\cite{IPC4_Domains}.}.

    \item The {\small \verb|data-network|} domain tackles distributed computing related problems. In a given network of servers, each server can produce data by processing some existing data and sends this to other servers on the network. The goal is to process data dispatched across servers with variable hardware and connection capabilities while minimizing the processing cost.

    \item Finally, the {\small \verb|Openstacks|} domain is based on the ``minimum maximum simultaneous open stacks'' combinatorial optimization problem. A manufacturer has orders for a combination of different products and can only make one product at a time, e.g., schedule order completion (NP-hard).

\end{itemize}

\begin{figure*}[!htb]
\captionsetup[subfloat]{farskip=0pt,captionskip=0pt}
\centering

\subfloat[]{\includegraphics[width=.33\textwidth]{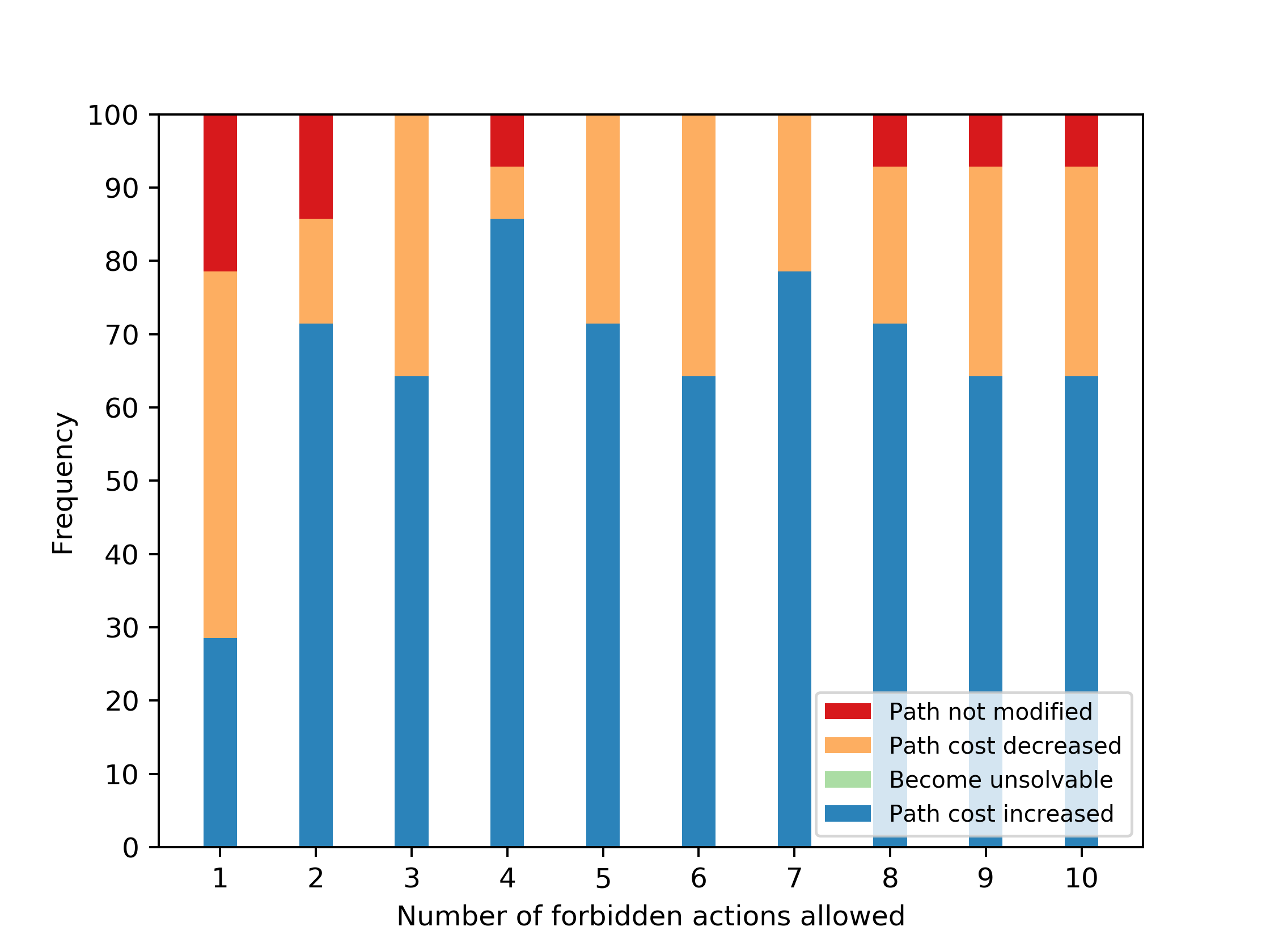}\label{FD_Bar_barman}}
\subfloat[]{\includegraphics[width=.33\textwidth]{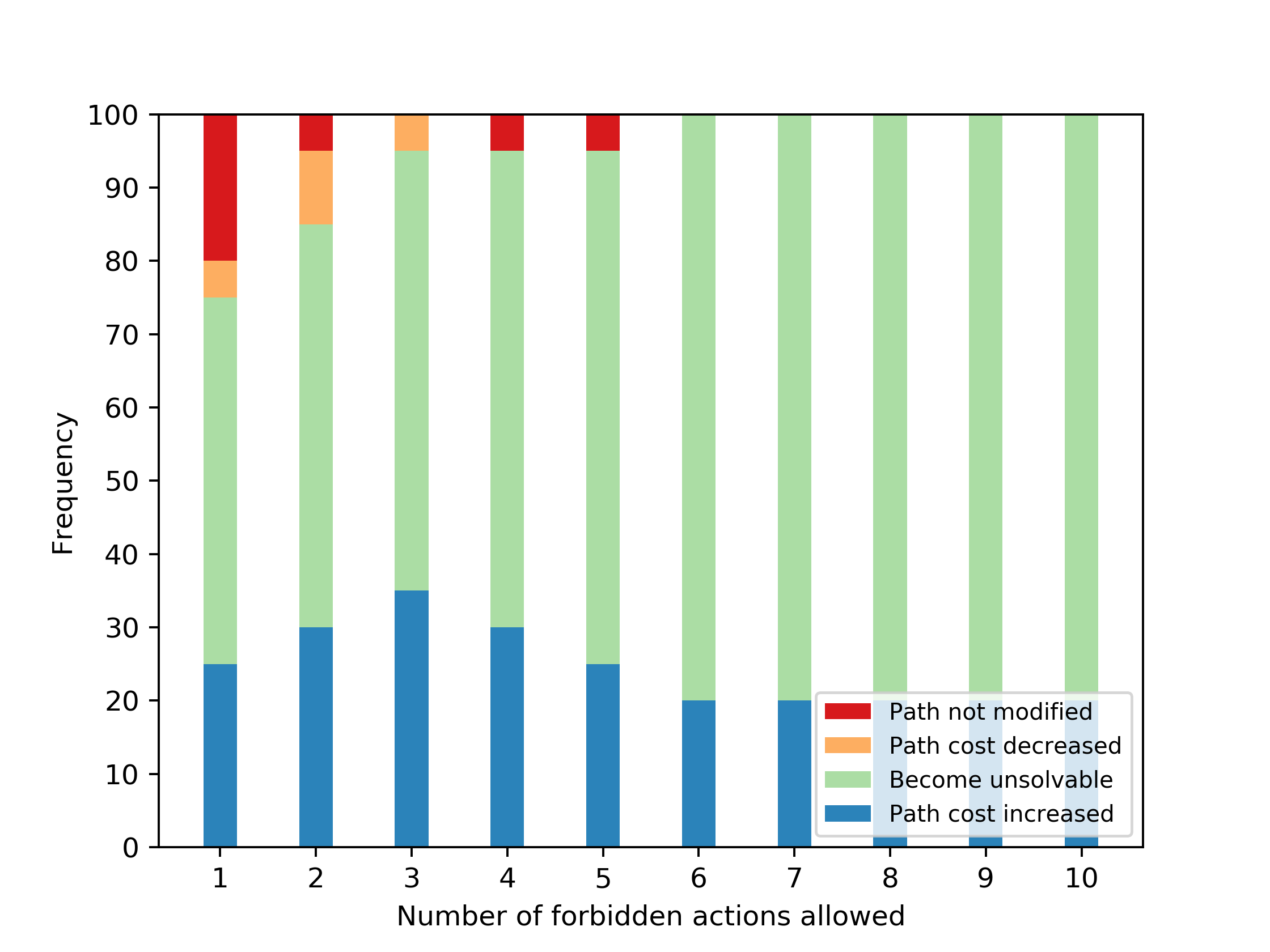}\label{FD_Bar_floortile}}
\subfloat[]{\includegraphics[width=.33\textwidth]{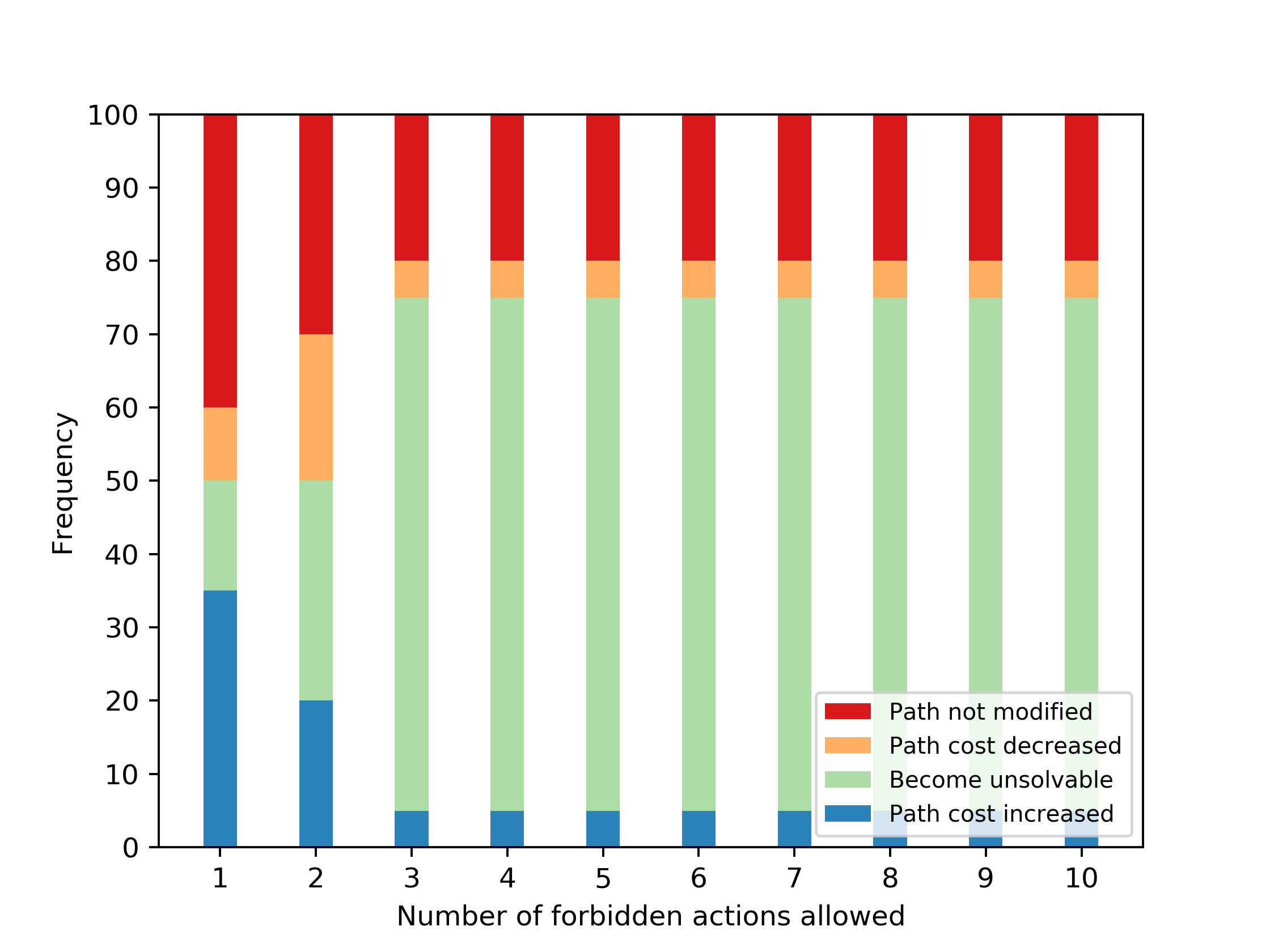}\label{FD_Bar_hiking}}

\subfloat[]{\includegraphics[width=.33\textwidth]{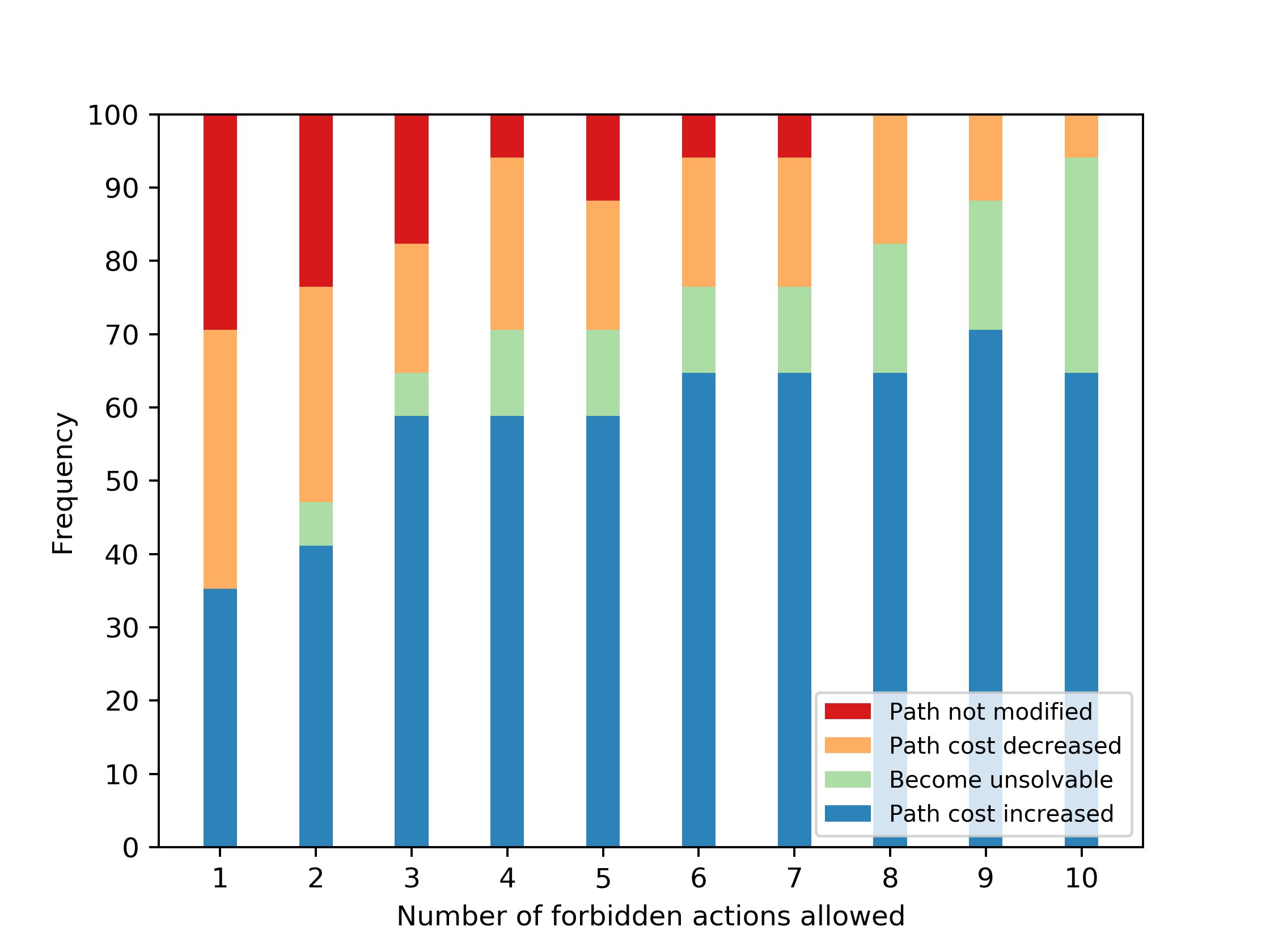}\label{FD_Bar_tetris}}
\subfloat[]{\includegraphics[width=.33\textwidth]{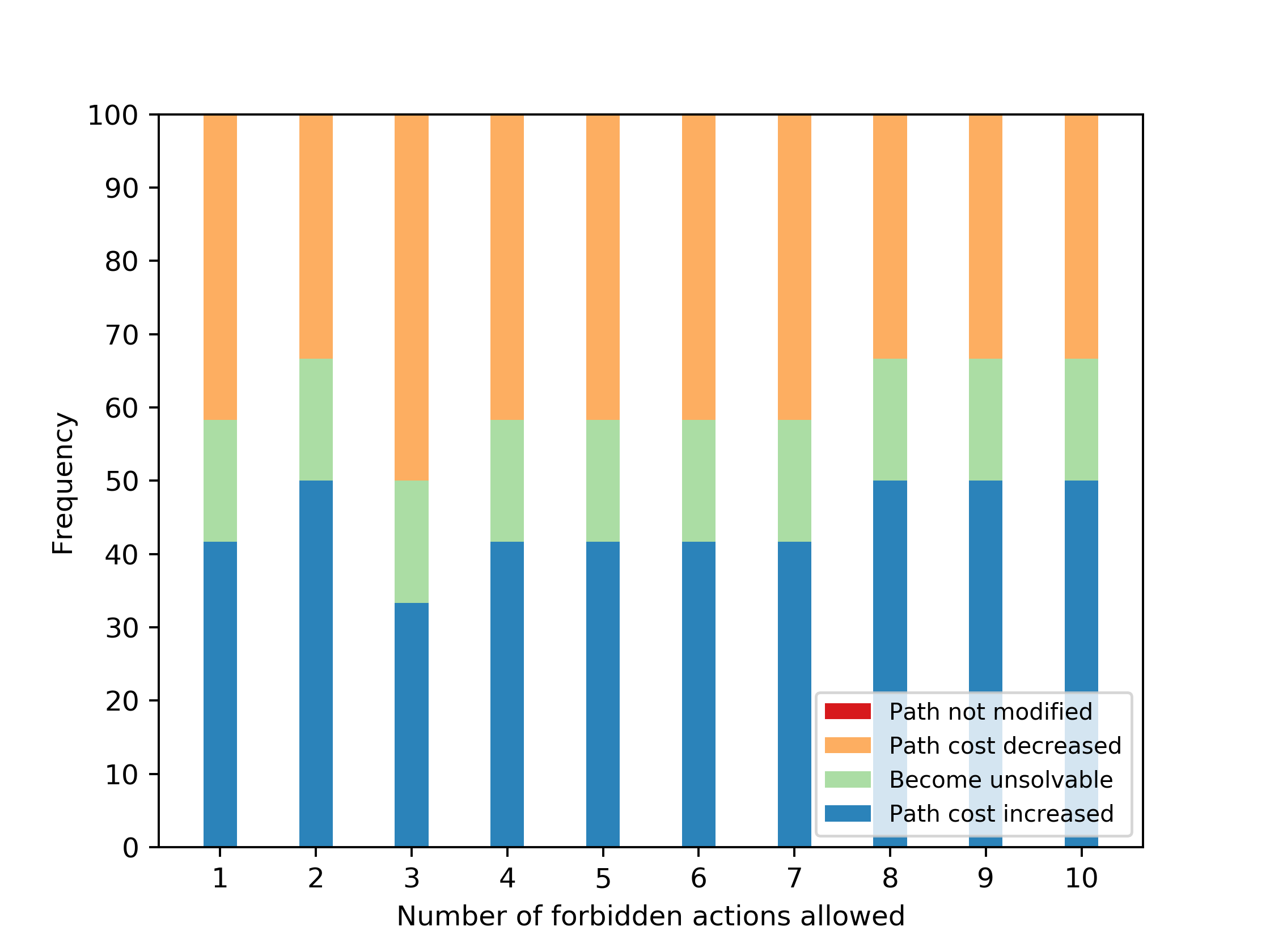}\label{FD_Bar_network}}

\cprotect\caption{Domains: (a) {\small \verb|barman|}, (b) {\small \verb|floortile|}, (c) {\small \verb|hiking|}, (d) {\small \verb|tetris|} and (e) {\small \verb|data-network|}. We benchmarked the window-heuristic using all the tasks available in the optimizing track. We show the success rate of the window-heuristic depending on the number of adversarial changes allowed (adversarial budget).}
\label{Evaluation_FD}
\end{figure*}

The results of our experiments are multifaceted.  First, the most readily perturbed domains (\ie most vulnerable to attack) were the ones where the goal cannot be divided into independent sub-goals. Assuming a planning instance can be divided into multiple independent sub-planning tasks with sub-goals, an agent can find a sub-plan to reach each sub-goal independently. The agent then assembles each sub-plan in an arbitrary order to create the final plan. An adversarial change is likely to perturb the sub-plan to achieve one of the sub-goals, but not the other sub-plans. For example, given a task in the {\small \verb|air cargo transportation|} domain with two packages to deliver ({\small \verb|c1|}, {\small \verb|c2|}). Each delivery is a sub-goal: we can deliver {\small \verb|c1|} first and then {\small \verb|c2|} or the opposite. An adversarial change only perturbs one of the sub-goals; the agent's planner just has to find another sub-goal-plan. This requires less work from the agent than finding another global plan. This is why the success rate of the {\small \verb|data-network|} domain remains around 50-60\%. On the other hand, tasks from the {\small \verb|floortile|} and {\small \verb|hiking|} domain cannot be divided into independent sub-goals and are consequently highly perturbed (Figure~\ref{Evaluation_FD}). Figure~\ref{Evaluation_FD} excludes the \verb|airport| and \verb|openstacks| domains as they become unsolvable early on in the number of adversarial changes allowed.

Because the cost of removing an action from the plan may have real cost (see Section~\ref{sec:realizing}), the number of changes may represent and important success parameter.  For example, in Figure~\ref{Evaluation_FD}, the {\small \verb|data-network|} domain does not seem to be a successful attack (58.32\% success rate with 4 grounded actions to remove). However, over few {\small \verb|data-network|} instances, an adversary would increase the cost on average by $73.83$ units with four adversarial changes.

\begin{anonsuppress}

Table~\ref{table:FD_avg_path_change} shows that, surprisingly, an adversary can also have a beneficial effect on the cost of the plan---the introduction of an adversary improves the plan by decreasing its cost. Intuitively, the introduction of the instance perturbation forces the planner (which does not guarantee optimality) to explore areas of the state space incorrectly pruned by an aggressive planning algorithm. When the agent explores these areas containing a shorter way to reach a goal state, the cost is decreased. Usually, we expect the agent's heuristic to guide the agent to these cost-effective areas. However, for some instances like the ones belonging to the {\small \verb|data-network|} domain, the search heuristic is sometimes ineffective because it produces plateaus. The agent usually breaks the ties by randomly selecting the next state, which makes its behavior unpredictable. Applying adversarial changes modifies the plateaus and may be favorable for the agent. This explains why the adversary has sometimes such a beneficial effect for instances in the {\small \verb|data-network|} domain.

\begin{table}[t]
\begin{center}
 \begin{tabular}{|c|c|c|c|c|c|c|c|c|c|c|}
 \hline
 \multicolumn{11}{|c|}{Number of adversarial changes allowed} \\
 \hline
 Domain & 1 & 2 & 3 & 4 & 5 & 6 & 7 & 8 & 9 & 10\\ [0.5ex]
 \hline\hline
Barman & 0.21 & 3.64 & 4.86 & 8.57 & 5.21 & 2.71 & 10.29 & 11.71 & 7.93 & 14.5 \\
Floortile & 2.6 & 3.95 & 4.65 & 5.05 & 4.55 & 4.05 & 4.05 & 3.55 & 3.35 & 3.35 \\
Hiking & 2.2 & 1.55 & 1.05 & 1.05 & 1.05 & 1.05 & 1.05 & 1.05 & 1.05 & 1.05 \\
Tetris & 9.18 & 10.47 & 15.00 & 16.47 & 14.18 & 15.00 & 16.24 & 16.53 & 18.65 & 23.29 \\
Airport & 4.72 & $\infty^*$\tnote{1} & $\infty$ & $\infty$ & $\infty$ & $\infty$ & $\infty$ & $\infty$ & $\infty$ & $\infty$\\
Openstacks & $\infty^*$\tnote{1} & $\infty$ & $\infty$ & $\infty$ & $\infty$ & $\infty$ & $\infty$ & $\infty$ & $\infty$ & $\infty$\\
Data-network & 56.75 & 87.67 & 34.42 & 73.83 & 72.83 & \color{ACMRed}-7.00 & 4.67 & 146.08 & 115.00 & 120.16\\
\hline
\end{tabular}
\caption{Average path cost impact. The tasks becoming unsolvable are ignored ($cost = +\infty$). In red, the adversary decreases the cost of the plan (on average). Entries marked with $^*$ mean up to that point all problems benchmarked become unsolvable, the plans have infinite cost and the average cost change is infinite. }

\label{table:FD_avg_path_change}

\end{center}
\end{table}
\end{anonsuppress}

\vspace{3pt}\noindent\underline{\textbf{Takeaways}} - We find that an adversary can be successful without knowing the agent's planner. The success rate of the adversary decreases if the tasks perturbed can be divided into sub-tasks with sub-goals. Overall, two adversarial changes are generally sufficient to efficiently perturb an agent's task.

\section{Realizing Attacks}
\label{sec:realizing}

One of the key questions one might ask is how an adversary can practically alter the plan instance.  Here we highlight the results of our ongoing survey of the security of real-world planning systems and demonstrate scenarios in which planning instances can be (and have been) targeted by adversaries.

\vspace{3pt}
\noindent
{\bf Transportation systems} - Next-generation transportation systems use vehicle area networks to exchange motion, location and hazard information (e.g., V2X~\cite{J29451}).  These vehicle to vehicle messages are used by internal planning systems in autonomous cars to determine how to make local and global decisions~\cite{zzwzc18,bt15}.  However, such messages can simply misreport the state of the environment, therein allowing selfish behavior~\cite{petit}.  Here, the misreports will alter the maneuver (action) space of the receiving victim cars, and therein alter the planning inputs as posited throughout.  Anecdotally, a recent low-tech attack on route planning was demonstrated by a performance artist Simon Weckert, who created a virtual traffic jam by carrying 99 phones on an otherwise empty street.  This fake ``congestion'' (perturbation of the plan space) was avoided by mapping software of those in the nearby area\cite{cox_how_2020}.

\vspace{3pt}
\noindent
{\bf Motion planning} - Motion planning in vision-based robotic systems is used at multi-scales to plan for the movement or manipulation of objects within the environment~\cite{motionplan98}.  However, it is known that vision systems used in robotics can be profoundly affected by changes in light.  In particular darkness, shadows and reflection, possibly caused by an adversary, can inhibit the scene or object interpretation/perception which can vastly alter the plan space~\cite{roboticslight99}.

\vspace{3pt}
\noindent
{\bf Chemical manufacturing} - Similar to other industrial manufacturing planning systems, chemical production scheduling is the process of scheduling, delivering and retrieving chemical components through a plant.  Such planning is highly dependent on the correct understanding of the available resource inventory and equipment states as recorded in plant databases~\cite{Kidam2013AnalysisOE}.  An adversary who is able to compromise the DB server (through techniques such as phishing, APT, or exploiting host vulnerabilities) can alter the database to
manipulate the planning of the production schedule.

\vspace{3pt}
\noindent
{\bf Data center/cloud management} - Data centers migrate virtual machines and containers to, among other goals, balance load, reduce resource usage, and provide isolation for sensitive computation.  These migrations are most often coordinated using a discrete or continuous resource planner~\cite{zs16}.  Any adversary who is able to occupy a VM host or surrounding infrastructure and generate network and/or computational load will change the resource signature and alter the data center planning instance.

\begin{anonsuppress}
\section{Related Work}
\label{RelatedWork}
The inception of planning began with breadth-first search, designed by Moore et al. in 1959~\cite{moore} to solve mazes. Later, Heuristic search was introduced in 1965 to better navigate the search space by using a function to estimate the distance from the current position to the goal~\cite{newell1965search}. With the creation of A*~\cite{hart1968formal} and the development of the STRIPS language~\cite{STRIPS} (which introduced planning formalism), state-space search arose. In 1986, Pednault et al. created the \textit{ADL} language~\cite{pednault1989adl} with looser requirements for problem formulation than STRIPS, which enabled state-space search algorithms to tackle even more realistic problems.



Also falling within the domain of AI, machine learning algorithms have been similarly shown to be vulnerable to malicious inputs~\cite{AdversarialExamples, carlini2017towards,Papernot}.  Here several works have demonstrated how slight but carefully calculated perturbations in an input (called \textit{adversarial examples}) could control deep learning systems. This emerging area of research inspired us to investigate the impact of introducing an adversary to deterministic planning problems. Indeed, our research demonstrated similar phenomena: small changes to the planning instance can have drastic consequences on the output of a planner.

Other past, related efforts have explored the use of machine learning for path-finding. Xiang et al.~\cite{QLearningPathFinding} found adversarial examples on a reinforcement learning technique, as applied to path-finding. They performed similar experiments on mazes: adding walls on the way of the agent. Their agent used a Q-learning algorithm~\cite{Qlearning}---a reinforcement learning technique---to move to the goal.  Here, they focused on general mazes and did not employ any notion of windows as we explored throughout. They focused only on path-planning, while we create a general method that can be applied to any planning instance that can be expressed in the \textit{PDDL} language~\cite{PDDL2.1} (or a subset of it, \eg STRIPS).

Thematically close but with different goals, Culberson and Schaeffer~\cite{patternDB} introduced pattern databases as an optimization for planners. Pattern databases store the exact number of moves required to solve various sub-goals of a planning instance. This method enhances single-agent search by reducing the total number of nodes explored. Similar to our work, the algorithm tries to match patterns from a table (\ie graph isomorphism) during a search through the state space.
\end{anonsuppress}



\section{Conclusion} \label{Conclusion}

This paper has explored adversarial capabilities in planning systems. We introduced an adversarial heuristic and algorithm for an adversary to identify malicious modifications to the environment that disrupt the plan to induce high cost or prevent the goal state from being reached. This algorithm can be adapted to any kind of deterministic, single-goal planning problem online or offline and scale with the size of the planning instance. For some domains the approach is successful in 60\% to 95\% instances if the adversary knows the desired goal and state of the agent.

In future work, we plan to explore defenses and measures of robustness. We will also explore more complicated planning domains (\eg multiple goals, irreversible actions, etc.) and methods for training adversarial planning heuristics.  As well as the practical impacts of manipulated planners {\it in situ}. By experimenting and measuring the real-world impacts of manipulated planners, we can understand the methods and degree to which planning system are vulnerable in the wild.

\bibliography{tist-adplan}

\begin{thebibliography}{10}

\bibitem{DARPA}
Anthony Stentz and Martial Hebert.
\newblock A complete navigation system for goal acquisition in unknown
  environments.
\newblock {\em Autonomous Robots}, 2(2):127--145, Jun 1995.

\bibitem{natural_language}
Alexander Koller and J{\"o}rg Hoffmann.
\newblock Waking up a sleeping rabbit: On natural-language sentence generation
  with {FF}.
\newblock In {\em ICAPS}, 2010.

\bibitem{smart_greenhouses}
Malte Helmert and Hauke Lasinger.
\newblock The scanalyzer domain: Greenhouse logistics as a planning problem.
\newblock In {\em Proceedings of the 20th International Conference on Automated
  Planning and Scheduling, {ICAPS} 2010, Toronto, Ontario, Canada, May 12-16,
  2010}, pages 234--237, 2010.

\bibitem{bsnw80}
William~I. Bullers, Shimon~Y. Nof, and Andrew~B. Whinston.
\newblock Artificial intelligence in manufacturing planning and control.
\newblock {\em A I I E Transactions}, 12(4):351--363, 1980.

\bibitem{cyberPlanning}
Mark~S Boddy, Johnathan Gohde, Thomas Haigh, and Steven~A Harp.
\newblock Course of action generation for cyber security using classical
  planning.
\newblock In {\em ICAPS}, pages 12--21, 2005.

\bibitem{D*Lite}
Sven Koenig and Maxim Likhachev.
\newblock D*lite.
\newblock In {\em Proceedings of the Eighteenth National Conference on
  Artificial Intelligence and Fourteenth Conference on Innovative Applications
  of Artificial Intelligence, July 28 - August 1, 2002, Edmonton, Alberta,
  Canada.}, pages 476--483, 2002.

\bibitem{bt15}
S.~{Behere} and M.~{Torngren}.
\newblock A functional architecture for autonomous driving.
\newblock In {\em 2015 First International Workshop on Automotive Software
  Architecture (WASA)}, pages 3--10, 2015.

\bibitem{zzwzc18}
W.~{Zong}, C.~{Zhang}, Z.~{Wang}, J.~{Zhu}, and Q.~{Chen}.
\newblock Architecture design and implementation of an autonomous vehicle.
\newblock {\em IEEE Access}, 6:21956--21970, 2018.

\bibitem{sar18}
Wilko Schwarting, Javier Alonso-Mora, and Daniela Rus.
\newblock Planning and decision-making for autonomous vehicles.
\newblock {\em Annual Review of Control, Robotics, and Autonomous Systems}, 1,
  may 2018.

\bibitem{tw16}
V.~L.~L. {Thing} and J.~{Wu}.
\newblock Autonomous vehicle security: A taxonomy of attacks and defences.
\newblock In {\em 2016 IEEE International Conference on Internet of Things
  (iThings) and IEEE Green Computing and Communications (GreenCom) and IEEE
  Cyber, Physical and Social Computing (CPSCom) and IEEE Smart Data
  (SmartData)}, pages 164--170, Dec 2016.

\bibitem{arc+15}
M.~{Amoozadeh}, A.~{Raghuramu}, C.~{Chuah}, D.~{Ghosal}, H.~M. {Zhang},
  J.~{Rowe}, and K.~{Levitt}.
\newblock Security vulnerabilities of connected vehicle streams and their
  impact on cooperative driving.
\newblock {\em IEEE Communications Magazine}, 53(6):126--132, June 2015.

\bibitem{carlini2017towards}
Nicholas Carlini and David Wagner.
\newblock Towards evaluating the robustness of neural networks.
\newblock In {\em 2017 IEEE Symposium on Security and Privacy (SP)}, pages
  39--57. IEEE, 2017.

\bibitem{Papernot}
Nicolas Papernot, Patrick~D. McDaniel, Somesh Jha, Matt Fredrikson, Z.~Berkay
  Celik, and Ananthram Swami.
\newblock The limitations of deep learning in adversarial settings.
\newblock {\em 2016 IEEE European Symposium on Security and Privacy
  (EuroS\&P)}, pages 372--387, 2016.

\bibitem{kurakin2016adversarial}
Alexey Kurakin, Ian Goodfellow, and Samy Bengio.
\newblock Adversarial machine learning at scale.
\newblock {\em arXiv preprint arXiv:1611.01236}, 2016.

\bibitem{chen2018angora}
Peng Chen and Hao Chen.
\newblock Angora: Efficient fuzzing by principled search.
\newblock In {\em 2018 IEEE Symposium on Security and Privacy (SP)}, pages
  711--725. IEEE, 2018.

\bibitem{aschermann2019redqueen}
Cornelius Aschermann, Sergej Schumilo, Tim Blazytko, Robert Gawlik, and
  Thorsten Holz.
\newblock Redqueen: Fuzzing with input-to-state correspondence.
\newblock In {\em NDSS}, volume~19, pages 1--15, 2019.

\bibitem{pcz+16}
Brian Paden, Michal {\v C}{\'a}p, Sze~Zheng Yong, Dmitry Yershov, and Emilio
  Frazzoli.
\newblock A survey of motion planning and control techniques for self-driving
  urban vehicles.
\newblock {\em IEEE Transactions on Intelligent Vehicles}, 1, April 2016.

\bibitem{Kidam2013AnalysisOE}
Kamarizan Kidam and Markku Hurme.
\newblock Analysis of equipment failures as contributors to chemical process
  accidents.
\newblock {\em Process Safety and Environmental Protection}, 91(1):61 -- 78,
  2013.

\bibitem{ws19}
Robert Wall and Alison Slider.
\newblock "u.s. airlines report delays caused by system fault: Faa said it was
  aware that several airlines were experiencing issues with a flight-planning
  program".
\newblock {\em The Wall Street Journal}, April 2019.

\bibitem{FastDownward}
Malte Helmert.
\newblock The fast downward planning system.
\newblock {\em Journal of Artificial Intelligence Research}, 26:191--246, 2006.

\bibitem{Sokoban_Domain_Specific}
Andreas Junghanns and Jonathan Schaeffer.
\newblock Domain-dependent single-agent search enhancements.
\newblock In {\em Proceedings of the 16th International Joint Conference on
  Artifical Intelligence - Volume 1}, IJCAI'99, pages 570--575, San Francisco,
  CA, USA, 1999. Morgan Kaufmann Publishers Inc.

\bibitem{Russell}
Stuart Russell and Peter Norvig.
\newblock {\em Artificial Intelligence: A Modern Approach}.
\newblock Prentice Hall Press, Upper Saddle River, NJ, USA, 3rd edition, 2009.

\bibitem{GraphPlan}
Avrim~L. Blum and Merrick~L. Furst.
\newblock Fast planning through planning graph analysis.
\newblock {\em Artificial Intelligence}, 90(1):281 -- 300, 1997.

\bibitem{blackboxSAT}
Henry Kautz and Bart Selman.
\newblock {BLACKBOX}: A new approach to the application of theorem proving to
  problem solving.
\newblock In {\em AIPS98 Workshop on Planning as Combinatorial Search}, volume
  58260, pages 58--60, 1998.

\bibitem{Bonet01planningas}
Blai Bonet and H{\'e}ctor Geffner.
\newblock Planning as heuristic search.
\newblock {\em Artificial Intelligence}, 129:5--33, 2001.

\bibitem{participatingPlanners}
Miles~Tracy Karen~Scarfone, Wayne~Jansen.
\newblock Description of participant planners of the deterministic track.
\newblock {\em The Eighth International Planning Competition}, June 2014.

\bibitem{TREX}
Conor McGann, Frederic Py, Kanna Rajan, Hans Thomas, Richard Henthorn, and Rob
  Mcewen.
\newblock {T-REX}: A model-based architecture for {AUV} control.
\newblock {\em The International Conference on Automated Planning and
  Scheduling}, 2007.

\bibitem{MinCut}
David Karger.
\newblock Global min-cuts in {RNC} and other ramifications of a simple mincut
  algorithm.
\newblock In {\em Proceedings of the Fourth Annual ACM-SIAM Symposium on
  Discrete Algorithms}, pages 21--30, 01 1993.

\bibitem{Complexity1}
Kutluhan Erol, Dana~S. Nau, and V.S. Subrahmanian.
\newblock Complexity, decidability and undecidability results for
  domain-independent planning.
\newblock {\em Artificial Intelligence}, 76(1):75 -- 88, 1995.
\newblock Planning and Scheduling.

\bibitem{Complexity2}
Tom Bylander.
\newblock The computational complexity of propositional {STRIPS} planning.
\newblock {\em Artificial Intelligence}, 69(1):165 -- 204, 1994.

\bibitem{patternDB}
Joseph~C Culberson and Jonathan Schaeffer.
\newblock Pattern databases.
\newblock {\em Computational Intelligence}, 14(3):318--334, 1998.

\bibitem{FFHeuristic}
J\"{o}rg Hoffmann and Bernhard Nebel.
\newblock The ff planning system: Fast plan generation through heuristic
  search.
\newblock {\em J. Artif. Int. Res.}, 14(1):253--302, May 2001.

\bibitem{CEAHeuristic}
Malte Helmert and Hector Geffner.
\newblock Unifying the causal graph and additive heuristics.
\newblock In {\em Proceedings of the Eighteenth International Conference on
  Automated Planning and Scheduling, {ICAPS} 2008, Sydney, Australia, September
  14-18, 2008}, pages 140--147, 2008.

\bibitem{IPC4_Domains}
Stefan Edelkamp, Roman Englert, J{\"o}rg Hoffmann, Frederico dos S.~Liporace,
  Sylvie Thi{\'e}baux, and Sebastian Tr{\"u}g.
\newblock Engineering benchmarks for planning: the domains used in the
  deterministic part of {IPC-4}.
\newblock {\em J. Artif. Intell. Res.}, 26:453--541, 2006.

\bibitem{J29451}
{\em On-Board System Requirements for V2V Safety Communications}, mar 2016.

\bibitem{petit}
J.~{Petit} and S.~E. {Shladover}.
\newblock Potential cyberattacks on automated vehicles.
\newblock {\em IEEE Transactions on Intelligent Transportation Systems},
  16(2):546--556, April 2015.

\bibitem{cox_how_2020}
Kate Cox.
\newblock How to virtually block a road: {Take} a walk with 99 phones, February
  2020.

\bibitem{motionplan98}
Kamal Gupta and Angel~P. Pobil.
\newblock {\em Practical Motion Planning in Robotics: Current Approaches and
  Future Directions}.
\newblock John Wiley \& Sons, Inc., USA, 1998.

\bibitem{roboticslight99}
John~M. Hollerbach, William~B. Thompson, and Peter Shirley.
\newblock The convergence of robotics, vision, and computer graphics for user
  interaction.
\newblock {\em The International Journal of Robotics Research},
  18(11):1088--1100, 1999.

\bibitem{zs16}
Zoha Usmani and Shailendra Singh.
\newblock A survey of virtual machine placement techniques in a cloud data
  center.
\newblock {\em Procedia Computer Science}, 78:491--498, 12 2016.

\bibitem{moore}
Edward~F. Moore.
\newblock The shortest path through a maze.
\newblock {\em Proceedings of the International Symposium on the Theory of
  Switching, Harvard University Press.}, pages 285--292, 1959.

\bibitem{newell1965search}
Allen Newell and George Ernst.
\newblock The search for generality.
\newblock In {\em Proc. IFIP Congress}, volume~65, pages 17--24, 1965.

\bibitem{hart1968formal}
Peter~E Hart, Nils~J Nilsson, and Bertram Raphael.
\newblock A formal basis for the heuristic determination of minimum cost paths.
\newblock {\em IEEE transactions on Systems Science and Cybernetics},
  4(2):100--107, 1968.

\bibitem{STRIPS}
Richard~E. Fikes and Nils~J. Nilsson.
\newblock {STRIPS}: A new approach to the application of theorem proving to
  problem solving.
\newblock {\em Artificial Intelligence}, 2(3):189 -- 208, 1971.

\bibitem{pednault1989adl}
Edwin~PD Pednault.
\newblock {ADL}: Exploring the middle ground between strips and the situation
  calculus.
\newblock {\em Kr}, 89:324--332, 1989.

\bibitem{AdversarialExamples}
Alexey Kurakin, Ian~J. Goodfellow, and Samy Bengio.
\newblock Adversarial examples in the physical world.
\newblock {\em CoRR}, abs/1607.02533, 2016.

\bibitem{QLearningPathFinding}
Y.~Xiang, W.~Niu, J.~Liu, T.~Chen, and Z.~Han.
\newblock A pca-based model to predict adversarial examples on q-learning of
  path finding.
\newblock In {\em 2018 IEEE Third International Conference on Data Science in
  Cyberspace (DSC)}, pages 773--780, June 2018.

\bibitem{Qlearning}
Christopher J. C.~H. Watkins and Peter Dayan.
\newblock Q-learning.
\newblock {\em Machine Learning}, 8(3):279--292, May 1992.

\bibitem{PDDL2.1}
Maria Fox and Derek Long.
\newblock {PDDL2.1:} an extension to {PDDL} for expressing temporal planning
  domains.
\newblock {\em CoRR}, abs/1106.4561, 2011.

\bibitem{MVAP}
Amotz Bar-Noy, Samir Khuller, and Baruch Schieber.
\newblock The complexity of finding most vital arcs and nodes.
\newblock Technical report, University of Maryland at College Park, 1995.
\newblock Univ. of Maryland Institute for Advanced Computer Studies Report No.
  UMIACS-TR-95-96.

\bibitem{DBLP:journals/corr/abs-1804-09155}
Cristina Bazgan, Till Fluschnik, Andr{\'{e}} Nichterlein, Rolf Niedermeier, and
  Maximilian Stahlberg.
\newblock A more fine-grained complexity analysis of finding the most vital
  edges for undirected shortest paths.
\newblock {\em CoRR}, abs/1804.09155, 2018.

\bibitem{DPP}
Peta Masters and Sebastian Sardina.
\newblock Deceptive path-planning.
\newblock In {\em Proceedings of the Twenty-Sixth International Joint
  Conference on Artificial Intelligence, {IJCAI-17}}, pages 4368--4375, 2017.

\bibitem{failureRecovery}
A.~E. Howe.
\newblock Improving the reliability of artificial intelligence planning systems
  by analyzing their failure recovery.
\newblock {\em IEEE Transactions on Knowledge \& Data Engineering}, 7:14--25,
  02 1995.

\bibitem{failureRecoveryDistributed}
Naveed Arshad, Dennis Heimbigner, and Alexander~L. Wolf.
\newblock A planning based approach to failure recovery in distributed systems.
\newblock In {\em Proceedings of the 1st ACM SIGSOFT Workshop on Self-managed
  Systems}, WOSS '04, pages 8--12, New York, NY, USA, 2004. ACM.

\end{thebibliography}
\bibliographystyle{unsrt}

\begin{anonsuppress}
\section{Acknowledgments}

The authors would like to thank Nicolas Papernot for his insightful comments and the parallels he pointed out with adversarial machine learning. We also would like to thank Rachel King, Blaine Hoak, and Berkay Celik for helping us during the writing and formatting phase. This material is based upon work supported by, or in part by, the National Science Foundation under Grant No. CNS-1805310 and Grant No. Grant No. CNS-1900873.
\end{anonsuppress}

\newpage
\appendix
\section{Sample task from the air cargo transportation domain}\label{Aircargodomain_example}

The {\small \verb|air cargo transportation|} domain specifies a class of task where cargos must be delivered to a destination airport minimizing the cost of the plan. The domain is described by three different operator {\small \verb|Load|}, {\small \verb|Unload|} and {\small \verb|Fly|} each with unit cost. The {\small \verb|Load|} operator is used to load a cargo in an airplane, {\small \verb|Unload|} is the inverse operator. Finally, {\small \verb|Fly|} is used to freight a cargo from one airport to another when previously loaded into a plane.
\begin{figure}[!h]
{\small
\begin{Verbatim}[frame=single]
(:action LOAD
  :parameters (?c - cargo ?p - plane
               ?a - airport)
  :precondition (and (At ?c ?a) (At ?p ?a))
  :effect (and (In ?c ?p) (not (At ?c ?a))))

(:action UNLOAD
  :parameters (?c - cargo ?p - plane
               ?a - airport)
  :precondition (and (In ?c ?p) (At ?p ?a))
  :effect (and (At ?c ?a) (not (In ?c ?p))))

(:action FLY
  :parameters (?p - plane ?from - airport
               ?to - airport)
  :precondition (and (At ?p ?from))
  :effect (and (At ?p ?to) (not (At ?p ?from))))
\end{Verbatim}
\cprotect\caption{The {\small \verb|air cargo transportation|} domain.}
}
\end{figure}

Now we define a sample task belonging to the domain: we give the initial state and a specification of the goal state. Figure~\ref{air_cargo_transportation_pb_sample} describes an initial state with two cargos, one at SFO and one at JFK. The goal is to switch the position of those using the two planes.
\begin{figure}[!h]
{\small
\begin{Verbatim}[frame=single]
(:objects
    p1 p2 - plane
    c1 c2 - cargo
    SFO JFK - airport)
(:init
    (At c1 SFO)
    (At c2 JFK)
    (At p1 SFO)
    (At p2 JFK))
(:goal
    (and (At c1 JFK) (At c2 SFO)))
\end{Verbatim}
\cprotect\caption{A problem from the {\small \verb|air cargo transportation|} domain.}
\label{air_cargo_transportation_pb_sample}
}
\end{figure}

\section{Finding adversarial examples is NP-hard}\label{adversarialNP_HARD}
Given a planning instance, we show that finding $k$ adversarial changes to increase the length of the optimal plan is an NP-Hard problem.
We define the Adversarial Change Problem (ADVCP) as follows:\\
\textbf{Input} - A planning instance $\mathcal{I}=(S_{init},S_{goal}, \mathcal{O})$ where $\mathcal{O}$ is a set of grounded operators with non-negative cost and an integer $k\in \mathbb{N}$.\\
\textbf{Output} - A set of $k$ grounded actions whose removal from $\mathcal{O}$ maximize the cost increase of the optimal plan---$k$ adversarial changes.\\
The decision problem associated with ADVCP (D-ADVCP) is the following. ``Given a planning instance $\mathcal{I}=(S_{init},S_{goal}, \mathcal{O})$ and two integers $(k,h)\in \mathbb{N}^2$, is there $k$ grounded actions whose removal from $\mathcal{O}$ makes the length of the optimal plan at least $h$''.
\\\\
We also introduce the Most Vital Arcs Problem (MVAP):\\
\textbf{Input} - A graph $G=(V,E)$ (directed or undirected), an integer $k\in \mathbb{N}$, and two nodes $(s,t)\in V^2$. Each edge in $e\in E$ has a non-negative cost of $c(e)$.\\
\textbf{Output} - A set of $k$ edges---arcs---whose removal maximize the length increase of the optimal path between $s$ and $t$ in $G$.

The decision problem associated with the MVAP (D-MVAP) is the following. ``Given $G=(V,E)$, $(k,h)\in \mathbb{N}^2$ and $(s,t)\in V^2$, is there $k$ edges whose removal makes the length of the shortest path from $s$ to $t$ at least $h$?''.
\\\\
We prove that ADVCP is an NP-Hard problem by reducing D-MVAP to D-ADVCP. D-MVAP is known to be NP-Hard~\cite{MVAP}~\cite{DBLP:journals/corr/abs-1804-09155}. We first introduce the polynomial time reduction between instances of the two decision problems. In a second time, we show that the answer for an instance of D-ADVCP is ``yes'' if and only if the answer of the corresponding instance of D-MVAP is ``yes''.

Suppose we are given an instance of the D-MVAP consisting of a graph $G=(V,E)$, $(k,h)\in \mathbb{N}^2$ two integers and $(s,t)\in V^2$ two nodes. Let $|V| = n$ and assume the nodes in $V$ are labelled $1,2,3,...,n$. We create the following equivalent D-ADVCP instance $\mathcal{I}$ (planning instance).
\begin{itemize}
    \setlength{\itemsep}{1pt}
    \setlength{\leftmargini}{1.3em}
    \item We define $S_{init}$ as the state with the following set of predicate: $\{${\small \verb|(Node, s)|}$\}$
    \item A state will be recognized as a goal state $S_{goal}$ if its predicates include the following set of predicates: $\{${\small \verb|(Node, t)|}$\}$
    \item If node $i$ is connected to $j$ with an edge in $G$, we add the following operator $o_{ij}$ to $\mathcal{O}$. The cost of this operator is set to the same cost as the $i$-to-$j$-edge cost.
    \begin{quote}
    \begin{Verbatim}[xleftmargin=-4em,fontsize=\small,frame=single]
    (:grounded-action Oij
      :precondition (Node, i)
      :effect (and (Node, j) (not (Node, i))))
    \end{Verbatim}
    \end{quote}
\end{itemize}

The resulting planning instance can be constructed in polynomial time in the size of $V$ and $E$. We now show that the answer for an instance of the D-ADVCP is ``yes'' if and only if the answer of the corresponding D-MVAP instance is ``yes''.

Assuming the answer is ``yes'' for a D-ADVCP instance $\mathcal{I}=(S_{init},S_{goal}, \mathcal{O})$ where $S_{init}=\{${\small \verb|(Node, s)|}$\}$ and $S_{goal}=\{${\small \verb|(Node, t)|}$\}$. Then removing $k$ grounded actions from $\mathcal{O}$ makes the length of the optimal plan at least $h$. We call $G_\mathcal{I}$ the graph given by the inverse reduction. $G_\mathcal{I}$ is exactly the state space graph generated by $\mathcal{I}$. The cost of $\mathcal{I}$'s optimal plan is the same as the cost of an optimal path from $S_{init}$ to $S_{goal}$ in $G_\mathcal{I}$. The graph $G_\mathcal{I}$, $(k,h)$ and $(S_{init},S_{goal})$: equivalent D-MVAP instance will be a ``yes''.

If $G=(V,E)$, $(k,h)\in \mathbb{N}^2$ and $(s,t)\in V^2$ (D-MVAP instance) is a ``yes'', there exist $k$ edges whose removal makes the length of the shortest path from $s$ to $t$ at least $h$. The equivalent planning instance $\mathcal{I}_G$ defined by the reduction has the same state space graph as $G$. Then, removing the $k$ equivalent grounded actions from $\mathcal{O}$ in $\mathcal{I}_G$ will increase the cost of the optimal plan by at least $h$.

Hence ADVCP is an NP-Hard problem. Note that we defined the ADVCP problem such that $\mathcal{I}=(S_{init},S_{goal}, \mathcal{O})$ would be given as an input. However, in real-life, the set $\mathcal{O}$ is not given, instead we give a set of non-grounded operators with a set of objects to ground them. Unfortunately, $\mathcal{O}$ can grow exponentially with the number of non-grounded operators and the number of objects. In the end, finding adversarial changes is finding an NP-hard problem's solution for an input with exponential size.

\section{Table generation parameters}\label{appdx:calib}

\begin{table}[h]
\centering
 \begin{tabular}{|c|c|c|c|c|}
 \hline
Threat model & Threshold & Window\, & Algorithm\, & $H_{adv}$\\ [0.5ex]
 \hline\hline
AHI\, & 0 & 231 & D*Lite & $h_{Euclidean}$\\
Agent's heuristic & 0 & 231 & D*Lite & $h_{Euclidean}$\\
Black-box & 10 & 45 & D*Lite & $h_{Manhattan}$ \\
\hline
\end{tabular}

\caption{The table generation process' settings for the different scenarios in the Maze domain. The window corresponds to the number of windows kept in the table, the algorithm is the one used by the adversary, and the AHI is the agent's heuristic and informed.}
\label{table:Maze_generation}
\end{table}

\section{Defenses} \label{Discussion}



It is natural to ask what defenses would mitigate this kind of adversarial planning. Consider that the agent will always follow what seems to be the shortest way to reach the goal. If we were to implement a new kind of planner resisting adversarial changes, we would face two contradictory incentives. (1) Try to find the shortest plan to the goal state and the plan becomes predictable or (2) find a less predictable plan that can be worse in terms of resources than the initial one. The first trade-off is the one followed by most planners and we showed it was sensible to an adversary. The second choice is exactly what an adversary wants because the plan is in the end worse than the initial one. This tension between these two contradictory goals makes it hard to come up with a defense against adversarial changes.

Deception and secrecy might be the only way to prevent an adversary from interfering the plan. In~\cite{DPP}, researchers worked on deceptive path planning: finding a path such that an observer cannot determine the goal the agent wants to reach until the last steps. Without a clear description of the goal state, an adversary is unable to predict the agent's plan. 

If an adversary still succeeds in finding an effective adversarial change, a way to limit its impact is to use failure recovery techniques. Howe~\cite{failureRecovery} studied plan resilience and error recovery at planning and execution time. Arshad et al.~\cite{failureRecoveryDistributed} investigated failure recovery for distributed system using planning. Their approach automates failure recovery by defining an acceptable recovered state as a goal. Then their system runs a planner to get from the current failure state to the recovered state. However that planner can also be vulnerable to adversarial changes.


\end{document}